\documentclass[12pt]{article}
\usepackage{amscd,amssymb,amsmath,latexsym,enumerate}
\usepackage[mathscr]{euscript}
\usepackage{epsfig}
\usepackage{fancybox}
\usepackage{verbatim}

\usepackage[normalem]{ulem}
\usepackage{multirow}
\usepackage{stackengine}

\usepackage{color}

\title{Finite volume calculation of $K$-theory invariants}

\author{Terry A. Loring$^1$ and Hermann Schulz-Baldes$^2$
\\
\\
\\
{\small $^1$Department of Mathematics and Statistics, University of New Mexico, USA}
\\
{\small $^2$Department Mathematik, Friedrich-Alexander-Universit\"at Erlangen-N\"urnberg, Germany}
}

\date{ }

\newtheorem{theo}{Theorem}
\newtheorem{defini}{Definition}
\newtheorem{proposi}{Proposition}
\newtheorem{lemma}{Lemma}
\newtheorem{coro}{Corollary}

\newcommand{\CM}{{\mathbb C}}

\newcommand{\RM}{{\mathbb R}}

\newcommand{\TM}{{\mathbb T}}
\newcommand{\ZM}{{\mathbb Z}}

\newcommand{\DM}{{\mathbb D}}

\newcommand{\Aa}{{\cal A}}
\newcommand{\Bb}{{\cal B}}

\newcommand{\Vv}{{\cal V}}

\newcommand{\Qq}{{\cal Q}}
\newcommand{\Kk}{{\cal K}}
\newcommand{\Hh}{{\cal H}}

\newcommand{\ii}{\imath} 

\newcommand{\one}{{\bf 1}}

\newcommand{\Tr}{\mbox{\rm Tr}}

\newcommand{\SF}{{\rm Sf}} 
\newcommand{\Pf}{{\rm Pf}} 
\newcommand{\Ch}{{\rm Ch}} 
\newcommand{\Ind}{{\rm Ind}} 
\newcommand{\Ker}{{\rm Ker}} 
 
\newcommand{\sgn}{{\rm sgn}} 
\newcommand{\Sig}{{\rm Sig}} 
\newcommand{\diag}{{\rm diag}}

\newcommand{\Alge}{{\cal B}}
\newcommand{\SL}{L} 
\newcommand{\WSL}{W} 
\newcommand{\VSL}{V} 
\newcommand{\Tune}{\kappa} 
\newcommand{\TapeP}{P} 
\newcommand{\TapeN}{N} 
\newcommand{\TapeG}{G_\rho} 
\newcommand{\Radius}{\rho} 
\newcommand{\TapePFunc}{p} 
\newcommand{\TapeNFunc}{n} 
\newcommand{\Perturb}{{\cal V}}
\newcommand{\sign}{s} 
\newcommand{\DiracP}{D'}

\begin{document}

\maketitle

\begin{abstract}
Odd index pairings of $K_1$-group elements with Fredholm modules are of relevance in index theory, differential geometry and applications such as to topological insulators. For the concrete setting of operators on a Hilbert space over a lattice, it is shown how to calculate the resulting index as the signature of a suitably constructed finite-dimensional matrix, more precisely the finite volume restriction of what we call the spectral localizer.  In presence of real symmetries, secondary $\ZM_2$-invariants can be obtained as the sign of the Pfaffian of the spectral localizer. These results reconcile two complementary approaches to invariants of topological insulators.

\vspace{.1cm}



\end{abstract}

\vspace{.5cm}

\section{Overview and main results}


\subsection{Odd dimensional index pairings}
\label{sec-OddDimInd}

To start out, let us spell out an example of an invariant from classical differential topology which can be calculated by the tools described below. Suppose we are given a smooth function $k\in\TM^d\mapsto A(k)$ of complex invertible $N\times N$ matrices on the  torus $\TM^d$ of odd dimension $d$. An invariant of this function $A$ is the odd Chern number given by
\begin{equation}
\label{eq-ChernDef}
\Ch_d(A)
\;=\;
\frac{(\frac{1}{2}(d-1))!}{d !}
\;
\left(\frac{\ii}{2\pi}\right)^{\frac{d+1}{2}}\,
\int_{\TM^d}\Tr\left(\big ( A^{-1} {\bf d} A \big )^d \right)
\;,
\end{equation}
where ${\bf d}$ denotes exterior differentiation. Note that for $d=1$ and $N=1$ this is just the winding number of a complex-valued function and therefore $\Ch_d(A)$ is also called a generalized winding number. It can be interpreted as the result of a paring of the class in $K^1(\TM^d)$ specified by $A$ with a de Rham cohomology class and thus is indeed a homotopy invariant. The normalization constant is chosen such that $\Ch_d(A)\in\ZM$. Actually, it is possible to calculate $\Ch_d(A)$ as the index of a Fredholm operator in the following manner. Suppose we are given a faithful irreducible representation  of the complex Clifford algebra $\CM_d$ by selfadjoint matrices $\Gamma_1,\ldots,\Gamma_d$ on $\CM^N$ (possibly given only after augmenting $N$).  Consider the associated Dirac operator $D=\imath\sum_{j=1}^d\Gamma_j\,\partial_{k_j}$ on $L^2(\TM^d,\CM^N)$ as well as its positive spectral projection $\Pi=\chi(D\geq 0)$, also called the Hardy projection. Then viewing $A$ as a multiplication operator on $L^2(\TM^d,\CM^N)$, the operator $\Pi A\Pi+(\one-\Pi)$ is Fredholm and the following index theorem holds:
\begin{equation}
\label{eq-ClassIndex}
\Ch_d(A)
\;=\;
\Ind\big(\Pi A\Pi+(\one-\Pi)\big)
\;.
\end{equation}
For $d=1$ and $N=1$ this is historically the very first index theorem proved by Fritz Noether  in 1920 \cite{Noe}. The case of larger $N$ goes back to at least Gohberg and Krein \cite{GK}. For larger odd $d$, a proof is contained as a special case in \cite{PSB}, which also considers extension to non-commutative crossed product algebras, but there likely exist earlier contributions for $d\geq 3$. The particular form of the Fredholm operator $\Pi A\Pi+(\one-\Pi)$ on the r.h.s. of \eqref{eq-ClassIndex} always appears in index theorems, both in classical differential topology and in non-commutative geometry \cite{Con}. It is the main object of the analysis below.

\subsection{Aims of the paper}
\label{sec-Aims}

The main aim of this paper is to provide an alternative way to calculate the index in \eqref{eq-ClassIndex} as the signature of a suitably constructed finite dimensional matrix which we call the {\sl spectral localizer}. As will be discussed below, this makes the invariant calculable by numerical means in interesting applications, and, in particular, also when there is no classical differential calculus available so that \eqref{eq-ChernDef} fails and non-commutative analysis tools are needed. From an analytic perspective, this allows to calculate the index of the Fredholm operator $\Pi A\Pi+(\one-\Pi)$, an intrinsically infinite dimensional object, from finite dimensional analysis. 

\vspace{.2cm}

As already stressed in the title of the paper and in Section~\ref{sec-OddDimInd}, the above Fredholm operator stems from the pairing of a $K_1$-class with the Hardy projection of a Dirac operator. In the terminology of $K$-homology and non-commutative geometry, the latter fixes an unbounded $K$-cycle or an unbounded odd Fredholm module. In Section~\ref{sec-ImpSym} we will further implement symmetries  invoking a real structure and then the spectral localizer also allows to calculate parings of $KR$-group elements with $KR$-cycles, still as signature or as sign of the Pfaffian of the spectral localizer. 

\vspace{.2cm}

While all these abstract structures are in the background and actually tools from $K$-theory will be essential for our proof of the main result, it can and will be stated by only appealing to basic notions of functional analysis, see Section~\ref{sec-FiniteVolSpec}. We hope that this makes the result accessible to a wider mathematical audience and to users from the field of {\it numerical $K$-theory}. In Section~\ref{sec-Perspectives} we then give a complementary $K$-theoretic perspective on the main results.

\subsection{Construction of the spectral localizer}

To construct the spectral localizer and at the same time considerably enlarge the class of index pairings beyond the example in Section~\ref{sec-OddDimInd}, we use the discrete Fourier transform to pass from $L^2(\TM^d,\CM^N)$ to $\ell^2(\ZM^d,\CM^N)$. The (dual) Dirac operator then becomes
\begin{equation}
\label{eq-DiracDef}
D\;=\;
\sum_{j=1}^d\,\Gamma_j\,X_j
\;,
\end{equation}
where, as above, the faithful representation of the Clifford algebra acts on the matrix degrees of freedom only and $X_1,\ldots,X_d$ are the $d$ components of the selfadjoint commuting position operators on $\ell^2(\ZM^d)$ defined by $X_j|n\rangle=n_j|n\rangle$, where $n=(n_1,\ldots,n_d)\in\ZM^d$ and $|n\rangle\in\ell^2(\ZM^d)$ is the Dirac Bra-Ket notation for the unit vector localized at $n$. The Fourier transform of the multiplication operator by $k\in\TM^d\mapsto A(k)$ is a bounded invertible operator $A$ on $\ell^2(\ZM^d,\CM^N)$ given by a discrete convolution.  The differentiability of $k\in\TM^d\mapsto A(k)$ implies that
\begin{equation}
\label{eq-CommBound}
\|[D,A]\|
\;<\;\infty
\;.
\end{equation}
This means that the commutator $[D,A]$ extends to a bounded operator. The bound \eqref{eq-CommBound} is one of the crucial hypothesis below. It holds for a much wider class of invertible operators $A$ on $\ell^2(\ZM^d,\CM^N)$ than those obtained by Fourier transform of a differentiable multiplication operator. From now on we will work with general invertible operators $A$ on the Hilbert space $\ell^2(\ZM^d,\CM^N)$ satisfying \eqref{eq-CommBound}, which we also call a locality bound on $A$. One of the consequences of \eqref{eq-CommBound} is that the operator $\Pi A\Pi+(\one-\Pi)$ is Fredholm where still $\Pi=\chi(D\geq 0)$ is the Hardy projection ({\it e.g.} p.~462 in \cite{GVF}). 

\vspace{.2cm}

From $D$ and $A$ let us now build two self-adjoint operators $\DiracP $ and $H$ on $\ell^2(\ZM^d,\CM^{2N})$:
\begin{equation}
\label{eq-DefHam}
\DiracP 
\;=\;
\begin{pmatrix}
D & 0 \\ 0 & -D
\end{pmatrix}
\;,
\qquad
H
\;=\;
\begin{pmatrix}
0 & A \\ A^* & 0
\end{pmatrix}
\;.
\end{equation}
Then the spectral localizer $\SL_\Tune$ associated to $A$ and tuning parameter $\Tune>0$ is defined by
\begin{equation}
\label{eq-DefBott}
\SL_\Tune\;=\;\Tune\,\DiracP \,+\,H
\;=\;
\begin{pmatrix}
\Tune D & A \\ A^* & -\Tune D
\end{pmatrix}
\;.
\end{equation}
Just as $D$, also $\DiracP$ is an unbounded operator with discrete spectrum which is not invertible. A standard procedure to eliminate the kernel is to add a constant mass term, either to $D$ or in the off-diagonal entries of $\DiracP$. It will, however, be part of the main result below that also with the local and invertible off-diagonal entry $A$ the spectral localizer $\SL_\Tune$ has trivial kernel. Let us note that, if $A$ is invertible, then so is $H$ and the spectral gaps $g=\|A^{-1}\|^{-1}=\|H^{-1}\|^{-1}$ coincide.

\subsection{Localized index pairings}
\label{sec-FiniteVolSpec}

Clearly the spectral localizer $\SL_\Tune=\SL_\Tune^*$ is self-adjoint. As $\DiracP $ is unbounded and has discrete spectrum, the bounded operator $H$ will be viewed as a perturbation. This perturbation does modify the eigenvalues. While those of $\DiracP $ lie symmetrically around the origin, there may well be a spectral asymmetry for $\SL_\Tune$. The main result of this paper states that this spectral asymmetry can already be read off from finite volume approximants of $\SL_\Tune$ and that it is equal to the index of the Fredholm operator discussed above. This finite volume restriction is
\begin{equation}
\label{eq-DefBott2}
\SL_{\Tune,\Radius}\;=\;
\begin{pmatrix}
\kappa D_\Radius & A_\Radius \\ A_\Radius^* & -\kappa D_\Radius
\end{pmatrix}
\;,
\end{equation}
where $ D_\Radius$ and $A_\Radius$ are the (Dirichlet) restrictions to the finite dimensional Hilbert space $\ell^2(\DM_\Radius)\otimes\CM^{N}$ over the discrete ball $\DM_\Radius=\{x\in\ZM^d\,:\,\|x\|\leq \Radius\}$ of radius $\Radius>0$. Also the matrix $\SL_{\Tune,\Radius}$ will be referred to as the spectral localizer.

\begin{theo}
\label{theo-main}
Let $A$ be an invertible, bounded operator on $\ell^2(\ZM^d)\otimes\CM^N$,  satisfying \eqref{eq-CommBound}, with $d$ odd.  Provided that
\begin{equation}
\label{eq-MainCond1}
\|[D,A]\|\;\leq\;\frac{g^3}{18\,\|A\|\,\Tune}
\;,
\end{equation}
and
\begin{equation}
\label{eq-MainCond2}
\frac{2\,g}{\Tune}\;\leq\;\Radius
\;,
\end{equation}
the matrix $\SL_{\Tune,\Radius}$ is invertible and thus has a well-defined signature, which is given by
\begin{equation}
\label{eq-MainInd}
\frac{1}{2}\;\Sig(\SL_{\Tune,\Radius})
\;=\;
\Ind\big(\Pi A\Pi+(\one-\Pi)\big)
\;.
\end{equation}
\end{theo}

This result achieves our main goal to read off the topological information contained in the index from basic spectral data of the matrix $\SL_{\Tune,\Radius}$, justifying hence the name spectral localizer. Let us note that this localization is on the eigenvalues of $\SL_{\Tune,\Radius}$ with small absolute value. Actually, $\SL_{\Tune}$ has infinitely many positive and negative eigenvalues (and a compact resolvent) and the effect of $A$ is to determine the asymmetry of the low lying spectrum of $\SL_{\Tune}$, which can then already be read off from $\SL_{\Tune,\Radius}$. In our situation where $D$ is given by \eqref{eq-DiracDef}, this localization also takes place in the physical space  of the underlying lattice $\ZM^d$ so that it also makes sense to speak of spatial localization. This discussion also  justifies the following terminology:

\begin{defini}
\label{def-LocSig}
The half-signature on the l.h.s. of \eqref{eq-MainInd} is called the {\rm localized index pairing} of the invertible operator $A$ with the Fredholm module specified by $D$.
\end{defini}

This deviates from the first author's work \cite{Lor} where $\SL_{\Tune,\Radius}$ was called the Bott operator and its half-signature the Bott index. While there are some good reasons to include Bott in the name (see Section~\ref{sec-Perspectives}), these terms have in the meanwhile been used in numerous publications for a different object \cite{LH2,DAR,TR,AS}. To avoid future confusion and also because of the broader mathematical scope linked to $\SL_{\Tune,\Radius}$ (see also Section~\ref{sec-Perspectives}), we suggest using spectral localizer for $\SL_{\Tune,\Radius}$ as well as Definition~\ref{def-LocSig} in the future.

\vspace{.2cm}

The proof of Theorem~\ref{theo-main} will ultimately be given at the end of Section~\ref{sec-estimate}. Here is a way to apply the result. From $A$ one first infers $\|[D,A]\|$, $\|A\|$ and the gap $g=\|A^{-1}\|^{-1}$, then next choses $\Tune$ sufficiently small such that the first bound \eqref{eq-MainCond1} holds, and uses the second bound \eqref{eq-MainCond2} to determine the minimal system size $\Radius_0$. Then just remains to build the finite matrix $\SL_{\Tune,\Radius}$ as in \eqref{eq-DefBott} and calculate its signature. This signature is equal to the index for any $\Radius\geq \Radius_0$, also arbitrarily large. If $g$, $\|A\|$ and $\|[D,A]\|$ are of the order of unity, then one infers roughly $\Radius_0\approx 100$. Hence only relatively small matrix sizes are needed.  We further note that for a unitary $A$, one has $\|A\|=g=1$ so that the bounds in Theorem~\ref{theo-main} somewhat simplify. More comments on the numerical implementation are given in Section~\ref{sec-TopIns}.

\vspace{.2cm}

Let us also add a few words of caution by discussing situations where Theorem~\ref{theo-main} does {\em not} apply. Suppose $d=1$ and that $A$ is given by the right shift on $[-2\Radius,2\Radius]$ and the identity outside of  $[-2\Radius,2\Radius]$. Now the signature of $\SL_{\Tune,\Radius}$ is $1$, but on the other hand the Fredholm operator $\Pi A\Pi+\one-\Pi$ is a compact perturbation of the identity and thus has vanishing index. The problem is, of course, that the invertibility of $A$ is a global assumption which is violated due to the defect at $-2\Radius$. If one reestablishes the invertibility by using periodic boundary conditions so that $A$ consists of the cyclic shift on $[-2\Radius,2\Radius]$, then the added matrix element leads to a commutator $[X,A]$ of the order of $\Radius$, which according to \eqref{eq-MainCond1} and \eqref{eq-MainCond2} forces one to use considerably larger volumes for the finite volume calculation, which then leads indeed to a vanishing signature invariant.


\subsection{Connection with the $\eta$-invariant}

The $\eta$-invariant was introduced by Atiyah-Patodi-Singer \cite{APS} as a measure of the spectral asymmetry of an invertible self-adjoint operator $L=L^*$ on a Hilbert space under the condition that $L$ has compact resolvent with eigenvalues decaying sufficiently fast such that $|L|^{-s}$ is trace class for $s>0$ sufficiently large. Then first the $\eta$-function is defined by
\begin{equation}
\label{eq-etadef}
\eta_s(L)
\;=\;
\Tr(L|L|^{-s-1})
\;=\;
\sum_j\sgn(\lambda_j)\,|\lambda_j|^{-s}
\;,
\end{equation}
where $\lambda_j$ are the eigenvalues of $L$. The $\eta$-function has a meromorphic extension given by
\begin{equation}
\label{eq-etadef2}
\eta_s(L)
\;=\;
\frac{1}{\Gamma(\tfrac{s+1}{2})}
\int_0^\infty dt\;t^{\tfrac{s-1}{2}}\;\Tr(L\,e^{-tL^2})
\;.
\end{equation}
Whenever $\eta_s(L)$ is regular at $s=0$, one says that the $\eta$-invariant of $L$ is well-defined and given by
\begin{equation}
\label{eq-EtaDef}
\eta(L)
\;=\;
\eta_0(L)
\;=\;
\frac{1}{\sqrt{\pi}}
\int_0^\infty dt\;t^{-\tfrac{1}{2}}\;\Tr(L\,e^{-tL^2})
\;.
\end{equation}
As $L=L^*$ one then has $\eta(L)\in\RM$. Comparing with \eqref{eq-etadef}, one also sees that $\eta(L)$ can indeed be interpreted as a measure of the spectral asymmetry of the spectrum. If $L$ is a matrix, then clearly $\eta(L)$ exists and 
\begin{equation}
\label{eq-EtaSig}
\eta(L)
\;=\;
\Sig(L)
\;.
\end{equation}

\vspace{.2cm}

Getzler \cite{Get} pointed out that there is a close relation between the $\eta$-invariant, $\theta$-summable Fredholm modules and the JLO-cocycle \cite{JLO}. Further elements of this theory as well as an extension to the semifinite case were developed by Carey and Phillips \cite{CP}. For the setting described above, the following result is proved in Section~\ref{sec-EtaBott}. Roughly, it makes more precise in which sense the limit $\Radius\to\infty$ in Theorem~\ref{theo-main} may be taken.

\begin{theo}
\label{theo-eta}
Let $A$ be local on $\ell^2(\ZM,\CM)$ in the sense that \eqref{eq-CommBound} holds. Then the spectral localizer $\SL_\Tune$ defined in \eqref{eq-DefBott} has a well-defined $\eta$-invariant which is equal to twice the index in \eqref{eq-MainInd}. In particular, whenever the conditions \eqref{eq-MainCond1} and \eqref{eq-MainCond2} hold, 
$$
\eta(\SL_\Tune)
\;=\;
\Sig(\SL_{\Tune,\Radius})
\;.
$$
\end{theo}

As an application of Theorem~\ref{theo-eta}, we provide an alternative proof, for the case $d=1$, of Theorem~\ref{theo-main} in Section~\ref{sec-EtaBott}. It shows that the conditions \eqref{eq-MainCond1} and \eqref{eq-MainCond2} cannot be improved considerably. 


\subsection{Even dimensional pairings}
\label{sec-even}

Theorem~\ref{theo-main} only considers odd-dimensional systems leading to odd index pairings. This is all we actually prove in this paper, but as an outlook to future work let us state that the spectral localization technique also works for even index pairings. As an example of such a pairing, consider a projection $P$ on $\ell^2(\ZM^d,\CM^N)$ with $d$ even. The even-dimensional Dirac operator has a grading $\Gamma_{d+1}$ allowing  us to extract the Dirac phase $F$ as a unitary operator from $D|D|^{-1}=\binom{0\;\;F}{F^*\;0}$. Then the Fredholm operator $PFP+(\one-P)$ is the resulting even index pairing and its index is equal to the top Chern number of $P$ \cite{PSB}. On the other hand, one can construct an associated spectral localizer
$$
\SL_\Tune
\;=\;
\Tune\,D\,+\,(2P-\one)\Gamma_{d+1}
\;.
$$
In an upcoming publication we show that, if $\|[D,P]\|<\infty$, $\Tune$ is sufficiently small and $\Radius$ sufficiently large, the index of $PFP+(\one-P)$ is equal to the signature of the finite volume restriction $\SL_{\Tune,\Radius}$. It is then also possible to implement symmetries for such even index pairings, similar to what is done in Section~\ref{sec-ImpSym} for odd index pairings.

\subsection{Implementation of symmetries}
\label{sec-ImpSym}

Whenever the Hilbert space has a real structure, the invertible operator $A$ can be real, symmetric, quaternionic or antisymmetric and then specifies a class in $KR$-theory of a suitable operator algebra \cite{BL,GS}. Furthermore, also  the Dirac operator $D$ given in \eqref{eq-DiracDef} can have symmetry properties involving the  real structure, so that it defines a $KR$-cycle. In the spirit of the presentation above, we will {\it not} stress these abstract notions, but rather present here a hands-on approach showing how $\ZM_2$-invariants can be produced from the spectral localizer by using the sign of its Pfaffian, just as in the first author's earlier work \cite{Lor}. This will be established by appealing to the paper \cite{GS} by Grossmann and the second author which systematically analyzes the fate of the index pairings $T=\Pi A\Pi+(\one-\Pi)$ in the presence of real symmetries when the complex Hilbert space is equipped with a fixed real structure which we simply denote by a complex conjugation bar. It is shown in \cite{GS} that the irreducible representation $\Gamma_1,\ldots,\Gamma_d$ of the Clifford algebra can be chosen such that there exists a real unitary matrix $\Sigma$ on the representation space leading to
\begin{equation}
\label{tab-DiracSym}
\begin{tabular}{|c||c|c|c|c|}
\hline
$d\,$mod$\,8$ & $1$ & $3$  & $5$ & $7$ 
\\
\hline\hline
$\Sigma^*\,\overline{D}\,\Sigma=$ & $D$    & $-D$ & $D$ &  $-D$   
\\
$\Sigma^2=$ & $\one$ &   $-\one$  &  $-\one$  & $\one$  
\\
\hline
$\Sigma^*\,\overline{\Pi}\,\Sigma=$ & $\Pi$    & $\one-\Pi$ & $\Pi$ &  $\one-\Pi$   
\\
\hline
\end{tabular}
\end{equation}
As an example, let us consider $d=3$. Then $D=X_1\sigma_1+X_2\sigma_2+X_3\sigma_3$ where $\sigma_1,\sigma_2,\sigma_3$ are the Pauli matrices, namely $\sigma_1$ and $\sigma_3$ are real and $\sigma_2$ is purely imaginary. Hence here $\Sigma=\ii \sigma_2$.  The last line in \eqref{tab-DiracSym} follows from the second one as $\Pi=\chi(D>0)$ (some care is needed on the kernel of $D$, where $\Sigma$ has to be defined separately, see \cite{GS} for details). Hence $\Pi$ is respectively real, odd Lagrangian, quaternionic or even Lagrangian. 

\vspace{.2cm}

The other ingredient $A$ of the index pairing can be even symmetric, quaternionic, odd symmetric or real with respect to another real unitary symmetry operator $S$ which is supposed to be given: 
\begin{equation}
\label{tab-ASym}
\begin{tabular}{|c||c|c|c|c|}
\hline
$j\,$mod$\,8$ & $2$ & $4$  & $6$ & $8$ 
\\
\hline\hline
$S^*\,\overline{A}\,S=$ & $A^*$    & $A$ & $A^*$ &  $A$   
\\
$S^2=$ & $\one$ &   $-\one$  &  $-\one$  & $\one$  
\\
\hline
\end{tabular}
\end{equation}
Here the index $j$ is merely used for book keeping, in a way consistent with \cite{GS} which also specifies the associated Real $K$-theory. We, moreover, suppose that
$$
S\,\Sigma\;=\;\Sigma\,S\;,
\qquad
S\,D\;=\;D\,S
\;,
\qquad
\Sigma\,A\;=\;A\,\Sigma
\;.
$$
This is guaranteed if, {\it e.g.}~the representation space of $\Gamma_1,\ldots,\Gamma_d$ is tensored to the Hilbert space on which $A$ is acting, a situation that is given in the application to topological insulators (Section~\ref{sec-TopIns}). It is convenient to encode the tables \eqref{tab-DiracSym} and \eqref{tab-ASym} into four signs $\sign_D$, $\sign'_D$, $\sign_A$ and $\sign'_A$ by the following equations:
$$
\Sigma^*\,\overline{D}\,\Sigma\;=\;\sign_D\,D
\;,
\qquad
\Sigma^2\;=\;\sign'_D\,\one
\;,
\qquad
S^*\,\overline{A}\,S\;=\;A^{[\sign_A]}
\;,
\qquad
S^2\;=\;\sign'_A\,\one
\;,
$$
where $A^{[1]}=A$ and $A^{[-1]}=A^*$. Given a combination of these symmetries, the Noether index of $T=\Pi A\Pi+(\one-\Pi)$ can be forced to either be even or to vanish, and in the latter case it may nevertheless be possible that the parity of its nullity is a well-defined secondary $\ZM_2$-invariant: 
$$
\Ind_2(T)
\;=\;
\dim(\Ker(T))\,\mbox{\rm mod}\;2\;\in\;\ZM_2
\;.
$$
Theorem~1 in \cite{GS}, with the roles of $d$ and $j$ exchanged and shifted, states that index parings $T=\Pi A\Pi+(\one-\Pi)$ take the following values (using $\Ind$ and $\Ind_2$):
\begin{equation}
\label{tab-PairingVals}
\begin{tabular}{|c||c|c|c|c|}
\hline
$\Ind_{(2)}(T)$ & $j=2$ & $j=4$  & $j=6$ & $j=8$ 
\\
\hline\hline
$d=1$ & $0$    & $2\,\ZM$ & $\ZM_2$ &  $\ZM$   
\\
\hline
$d=3$ & $2\,\ZM$ & $\ZM_2$ &  $\ZM$    & $0$       
\\
\hline
$d=5$  & $\ZM_2$ &  $\ZM$    & $0$    & $2\,\ZM$
\\
\hline
$d=7$    & $\ZM$ & $0$    & $2\,\ZM$ & $\ZM_2$
\\
\hline
\end{tabular}
\end{equation}
This concludes the symmetry analysis of the index pairings on the r.h.s. of \eqref{eq-MainInd}. Now let us consider the l.h.s. and analyze the symmetries of the spectral localizers $\SL_\Tune$ given by \eqref{eq-DefBott}, as well as its finite volume restriction $\SL_{\Tune,\Radius}$. For that purpose, we diagonally extend $\Sigma$ and $S$ to $2\times 2$ matrices. Then
$$
(\Sigma S)^*\overline{\SL_\Tune}(\Sigma S)
\;=\;
\begin{pmatrix}
\Tune\,\Sigma^*\,\overline{D}\,\Sigma & S^*\,\overline{A}\,S
\\
(S^*\,\overline{A}\,S)^* & -\,\Tune\,\Sigma^*\,\overline{D}\,\Sigma
\end{pmatrix}
\;=\;
\begin{pmatrix}
\Tune\,\sign_D\,D & A^{[\sign_A]}
\\
(A^{[\sign_A]})^*  & -\,\Tune\,\sign_D\,D
\end{pmatrix}
\;.
$$
This can conveniently be rewritten using the Pauli matrices $\sigma_1=\binom{0\;1}{1\;0}$ and $\sigma_3=\binom{1\;\;\;0}{0\;-1}$. Setting $\sigma_j^{\{\sign\}}=(\sigma_j)^{\frac{1-\sign}{2}}$, one finds
$$
(\Sigma S)^*\overline{\SL_\Tune}(\Sigma S)
\,=\,
\sigma_1^{\{\sign_A\}}
\begin{pmatrix}
\Tune\,\sign_A\,\sign_D\,D & A
\\
A^*  & -\,\Tune\,\sign_A\,\sign_D\,D
\end{pmatrix}
\sigma_1^{\{\sign_A\}}
\,=\,
\sign_A\,\sign_D\,\sigma_1^{\{\sign_A\}} \sigma_3^{\{\sign_A\sign_D\}}\SL_\Tune\sigma_3^{\{\sign_A\sign_D\}}\sigma_1^{\{\sign_A\}}
\,.
$$
Introducing the real symmetry $R=\Sigma S\sigma_1^{\{\sign_A\}} \sigma_3^{\{\sign_A\sign_D\}}$ and signs $\sign_L$ and $\sign'_L$ by
$$
R^*\,\overline{\SL_\Tune}\,R\;=\;\sign_B\,\SL_\Tune
\;,
\qquad
R^2\;=\;\sign'_L\,
\one\;,
$$
one then has $\sign_L=\sign_A\,\sign_D$ and $\sign'_L=\sign'_D\sign'_A\,(\sign_A)^{\frac{\sign_D+1}{2}}$.  Inserting this into a table gives
\begin{equation}
\label{tab-BottVals}
\begin{tabular}{|c||c|c|c|c|}
\hline
$\sign_L=\;\;,\;\;\sign'_L=$& $j=2$ & $j=4$  & $j=6$ & $j=8$
\\
\hline\hline
$d=1$ & $-1\;,\;-1$    & $1\;,\;-1$ & $-1\;,\;1$ &  $1\;,\;1$   
\\
\hline
$d=3$ & $1\;,\;-1$    & $-1\;,\;1$ & $1\;,\;1$ &  $-1\;,\;-1$
\\
\hline
$d=5$  & $-1\;,\;1$    & $1\;,\;1$ & $-1\;,\;-1$ &  $1\;,\;-1$
\\
\hline
$d=7$    & $1\;,\;1$    & $-1\;,\;-1$ & $1\;,\;-1$ &  $-1\;,\;1$
\\
\hline
\end{tabular}
\end{equation}
As all real symmetry operators $\Sigma$, $S$ and $R$ are local (commute with the position operators), the symmetry properties of the finite volume approximations $\SL_{\Tune,\Radius}$ are the same as those of $\SL_\Tune$. The pattern of signs in \eqref{tab-BottVals} is the same as in \eqref{tab-PairingVals}, so it merely remains to understand why the four combination of signs $\sign_L$ and $\sign'_L$ imply that the invariant of $\SL_{\Tune,\Radius}$ takes the four different values appearing in \eqref{tab-PairingVals}. This is achieved in the following proposition.

\begin{proposi}
\label{prop-MatrixInvariants}
Let $L=L^*$ be an invertible complex matrix, and $R=\overline{R}$ a real unitary matrix of same size such that for two signs $\sign_L$ and $\sign'_L$
$$
R^*\,\overline{L}\,R\;=\;\sign_L\,L\;,
\qquad
R^2\;=\;\sign'_L\,\one
\;.
$$
\begin{itemize}

\item[{\rm (i)}] If $\sign_L=1$ and $\sign'_L=1$, then $\Sig(L)\in\ZM$ can take any integer value.

\item[{\rm (ii)}] If $\sign_L=1$ and $\sign'_L=-1$, then $\Sig(L)\in 2\,\ZM$ can take any even integer value.

\item[{\rm (iii)}] If $\sign_L=-1$ and $\sign'_L=1$, then $\Sig(L)=0$, but setting $M=R^{\frac{1}{2}}$ one obtains a real antisymmetric matrix $\ii MLM^*$ with invariant $\sgn(\Pf(\ii MLM^*))\in\ZM_2$.

\item[{\rm (iv)}] If $\sign_L=-1$ and $\sign'_L=-1$, then $\Sig(L)=0$.

\end{itemize}
\end{proposi}

\noindent {\bf Proof.} In (i) and (ii) the self-adjoint matrix $L$ is real and quaternionic respectively, implying the claim. The signature in (iii) and (iv) vanishes because $R^*\overline{L}R=-L$ and the signature is invariant under complex conjugation and matrix conjugation. In (iii), the first branch of the root is used so that the spectrum of $M$ is $\{\ii,1\}$. As $\overline{M}=M^*=M^{-1}$, the matrix $MLM^*$ is antisymmetric and selfadjoint so that $\ii MLM^*$ is real antisymmetric. It hence has a well-defined real Pfaffian $\Pf(\ii MLM^*)$, which cannot vanish because $L$ and $M$ are invertible. Let us note that choosing the other branch of the root leads to a different sign.
\hfill $\Box$

\vspace{.2cm}

In conclusion, provided the conditions \eqref{eq-MainCond1} and \eqref{eq-MainCond2} in Theorem~\ref{theo-main} hold, the eight integer valued invariants of the index pairing $\Pi A\Pi+(\one-\Pi)$ in \eqref{tab-PairingVals} can be calculated as the signature of the associated finite volume spectral localizer $\SL_{\Tune,\Radius}$. Furthermore, Proposition~\ref{prop-MatrixInvariants}(iii) suggests that the four $\ZM_2$-entries in \eqref{tab-PairingVals} can be calculated from the sign of the Pfaffian of the localizer. A formal proof of this fact is not given here. Let us note, however, that due to the homotopy invariance of both $\ZM_2$-invariants (by homotopies conserving the symmetries) it is sufficient to verify the equality on each connected component. In any case,  all the invariants extracted from $\SL_{\Tune,\Radius}$ by Proposition~\ref{prop-MatrixInvariants} are well-defined and are called {\sl Real localized index pairings}, similar as in Definition~\ref{def-LocSig}. Examples are given in the next section.

\subsection{Applications to topological insulators}
\label{sec-TopIns}

In this section, we indicate how the mathematical results of this paper can be applied to topological insulators. After comments on numerical implementation, we focus on a few physically relevant examples. A more detailed account with numerical results will be given elsewhere. For background information on topological insulators, we refer to \cite{RSFL,LH,Lor,GS,BCR,PSB,BKR}. 

\vspace{.2cm}

Theorem~\ref{theo-main} can immediately be applied to so-called chiral tight-binding Hamiltonians $H$ on $\ell^2(\ZM^d)\otimes\CM^{2N}$ of the form \eqref{eq-DefHam}. The chiral symmetry then reads $J^*HJ=-H$ where $J=\binom{\one\;\;\;0}{0\;-\one}$. The described system is then an insulator when $0$ is not in the spectrum of $H$, which is equivalent to the invertibility of $A$. Furthermore, the commutator bound \eqref{eq-CommBound} reflects the locality of $H$ and it generally holds for tight-binding Hamiltonians of physical interest. The index of $\Pi A\Pi+(\one-\Pi)$ is then \cite{PSB} precisely the strong invariant "higher winding number" used in the physics literature, {\it e.g.} \cite{RSFL}, which, when nonzero, makes the insulator into a topological one.  Theorem~\ref{theo-main} now states that it can be calculated as the localized index pairing. For $d=1$ and $d=3$, this was explicitly spelled out and numerically implemented in \cite{Lor} (and called the Bott index there, see the discussion in Section~\ref{sec-FiniteVolSpec}).

\vspace{.2cm}

Let us now advertise the advantages of the spectral localizer when it comes to the numerical calculation of topological indices. First of all, the definition of $\SL_{\Tune,\Radius}$ is directly given in terms of the Hamiltonian and does {\sl not} involve a spectral flattening as in many other numerical procedures. Secondly, the determination of the signature of $\SL_{\Tune,\Radius}$ by the block Chulesky algorithm is only polynomial in the system size and can hence be carried out very efficiently (note that the block Cholesky decomposition itself is not needed). To assure the validity of the result (and thus the good choices of $\Tune$ and $\Radius$), an easy test is to determine the gap size of $\SL_{\Tune,\Radius}$ by the inverse power method and verify that it is sufficiently large. Several further advantages explored in future work are the following. The spectral gap of $\SL_{\Tune,\Radius}$ still remains open in an Anderson localization regime and thus the method also applies in this regime. The particular choice of Dirichlet boundary conditions is completely irrelevant, merely locality of the boundary conditions is crucial. One can modify the definition of the spectral localizer to calculate weak topological phases and spin Chern numbers. Moreover, the local character of $\SL_{\Tune,\Radius}$ allows for the analysis of topologically inhomogeneous materials and hence changes in the quantum phase.  

\vspace{.2cm}

Next  we briefly mention other numerical approaches to disordered topological insulators.  In \cite{FHA} a scattering theory approach is implemented. The works \cite{SFP,Prod} successfully implement the non-commutative version of \eqref{eq-ChernDef} to calculate the invariants. Another attempt to localize topological information in physical space is \cite{BiR}. Finally the works \cite{LH2,DAR,TR,AS} use the Bott index.

\vspace{.2cm}

The above discussion only addressed complex topological insulators without particle-hole and time-reversal symmetry. If the chiral Hamiltonian has a supplementary real symmetry, we are in the framework of Section~\ref{sec-ImpSym}. The paper \cite{GS} systematically analyzes the symmetries of the Hamiltonian as well as the Dirac operators and shows how the periodic table of topological insulators \cite{RSFL} can be explained from an index theory point of view. Of the 64 real classes, the 16 cases of \eqref{tab-PairingVals} only correspond to the odd dimensional chiral systems. As the systems with integer invariants are dealt with directly by Theorem~\ref{theo-main}, let us highlight the two $\ZM_2$-invariants in low dimension. For $d=1$, if the SSH model has a even time-reversal symmetry and thus lies in Class DIII, one can calculate the $\ZM_2$-index from the spectral localizer by using Proposition~\ref{prop-MatrixInvariants}(iii). This agrees with Section~4.4 in \cite{Lor}. For $d=3$, the chiral Hamiltonian should have a supplementary odd time-reversal symmetry and thus lie in Class CII in order to have a non-trivial $\ZM_2$-invariant, calculable again by Proposition~\ref{prop-MatrixInvariants}(iii). 

\vspace{.2cm}

Let us point out that the tools of this paper only allow us to deal with chiral Hamiltonians and that this does not cover all $\ZM_2$-indices of interest. Important in dimension $d=1$ is the Class D with the so-called Kitaev chain as standard topologically non-trivial representative. Section 4.3 of \cite{Lor} states that the sign of the determinant of a spectral localizer gives the desired invariant in this case.  This invariant has been used for a numerical study of a model of a potential two-dimensional weak topological superconductor in Class D \cite{FPL}. To show that it coincides with the $\ZM_2$-index of \cite{GS} will be the another objective of a future publication.

\subsection{$K$-theoretic perspectives}
\label{sec-Perspectives}

The Dirac operator \eqref{eq-DiracDef} defines an unbounded odd Fredholm module (and thus a $K$-homology class) for the commutative C$^*$-algebra generated by the invertible operator $A$ on $\ell^2(\ZM^d,\CM^N)$. In the spirit of non-commutative geometry this specifies a spatial structure \cite{Con} and the finite volume restriction in \eqref{eq-DefBott2} is precisely w.r.t. to this notion of space. The particular form of the Hilbert space and $D$ was chosen to accommodate our applications to topological insulators, and because it naturally extends the introductory example of Section~\ref{sec-OddDimInd}. However, the technique of the spectral localizer applies to {\it any} odd index pairing constructed from an invertible operator $A$ and an unbounded odd Fredholm module for a C$^*$-algebra containing $A$. The $K$-theoretic proof given in the remainder of the paper goes through with minor modifications.



\vspace{.2cm}

The $K$-theoretic proof below is rooted in a general principle, namely that the index map of an odd-dimensional fuzzy sphere is an even-dimensional fuzzy sphere (and similarly, even-dimensional fuzzy spheres are mapped under the exponential map of $K$-theory to an odd-dimensional sphere, but this will not be used here). Before explaining this in some detail, let us state a definition and general fact (following readily from Proposition~\ref{prop-IndexMap} below):

\begin{defini}
\label{def-fuzzy}
Let $\Qq$ be a unital C$^*$-algebra. A fuzzy $d$-sphere of width $\delta<1$ is a collection of self-adjoints $Y_1,\ldots,Y_{d+1}\in\Qq$ with spectrum in $[-1,1]$ such that
$$
\Big\|\one -\sum_{j=1,\ldots,d+1} (Y_j)^2\Big\|\;<\;\delta\;,
\qquad
\|[Y_j,Y_i]\|\;<\;\delta
\;.
$$
\end{defini}

\begin{proposi}
\label{eq-OddSpheres}
Let $0 \rightarrow \Kk \,\hookrightarrow\,\Bb\,\rightarrow\, \Qq \rightarrow 0$ be a short exact sequence of C$^*$-algebras. If $d$ is odd, a fuzzy $d$-sphere in $\Qq$ specifies an element $[A]_1\in K_1(\Qq)$ via
$$
A\;=\;\sum_{j=1,\ldots,d} Y_j\,\Gamma_j\;+\;\imath\,Y_{d+1}
\;.
$$
If $B\in\Bb$ is a lift of $A$ and $B^*B=R^2$, the image of the index map in $K_0(\Kk)$ is given by
$$
\Ind[A]_1\;=\;
\left[
\begin{pmatrix}
2R^2-1 & 2(1-R^2)^{\frac{1}{2}}B
\\
2B^*(1-R^2)^{\frac{1}{2}} & -(2R^2-1)
\end{pmatrix}
\right]_0
\;.
$$
\end{proposi}

In our situation, $\Kk$ and $\Bb$ are the compact and bounded operators on $\Hh$ and $\Qq$ is the Calkin algebra. If $A$ is unitary,  the Fredholm operator $\Pi A\Pi+(\one-\Pi)$ is nothing by a unitary in the Calkin algebra and its real and imaginary part can hence also be seen as a (not so fuzzy) $1$-sphere. It is now shown in Theorem~\ref{theo-abstract} that $\Ind[A]_1$ can be understood as a fuzzy $2$-sphere in $\Kk$. The topological content of this fuzzy $2$-sphere is essentially the Bott projection, and it will be show that it can be read off of finite-volume approximations by the signature.

\vspace{.2cm}

Sections~\ref{sec-ImpSym} and parts of Section~\ref{sec-TopIns} alluded to notions of Real $K$-theory and Real $K$-homology. Despite considerable recent efforts \cite{BL,GS,BCR,BKR}, several aspects of the theory (in particular, the boundary maps) are not in a satisfactory state. Nevertheless, the cited papers allow the expert to readily read off the $KR$-homology classes of $D$ and $KR$-classes of $A$.

\section{The image of the index map}
\label{sec-ImageInd}

Let us begin by recalling  the definition of the $K$-groups of a C$^*$-algebra $\Aa$ which may be unital or non-unital. As in \cite{BL} or \cite[Section 4.2]{GS}, we prefer to work with self-adjoint unitaries rather than projections for the definition of $K_0(\Aa)$. The unitization $\Aa^+=\Aa\oplus\CM$ is equipped with the product  $(A,t)(B,s)=(AB+As+Bt,ts)$ and the adjunction $(A,t)^*=(A^*,\overline{t})$ as well as the natural C$^*$-norm $\|(A,t)\|=\max\{\|A\|,|t|\}$. The unit in $\Aa^+$ is $\one=(0,1)$. The unitization sits in an exact sequence of C$^*$-algebras $0 \rightarrow   \Aa  \overset{i}{\hookrightarrow} \Aa^+   
\overset{\rho}{\rightarrow} \CM \rightarrow  0$. A right inverse to $\rho$ is given by $i'(t)=(0,t)$, and the map $s=i'\circ \rho:\Aa^+\to\Aa^+$ extracts the scalar part. Set
\begin{equation}
\label{eq-V0def}
\Vv_0(\Aa)
\;=\;
\left\{
V\in \cup_{n\geq 1}M_{2n}(\Aa^+)\;:\;
V^*\,=\,V\;,
\;\;
V^2\,=\,\one\;,
\;\;
s(V)\sim_0 E_{2n}
\right\}
\;,
\end{equation}
where $M_{2n}(\Aa^+)$ are the $2n\times 2n$ matrices with entries in $\Aa^+$, and $s(V)\sim_0 E_{2n}$ forces the scalar part of $V$ to be homotopic to $E_{2n}=E_2^{\oplus^n}$ with $E_{2}=\binom{\one\;\;\;\;0\;}{0\;\;-\one}$.  An equivalence relation $\sim_0$ on $\Vv_0(\Aa)$ is defined by homotopy within the self-adjoint unitaries of fixed matrix size, complemented by 
\begin{equation}
\label{eq-equirel}
V\;\sim_0\;
\begin{pmatrix}
V & 0 \\
0 & E_2  
\end{pmatrix}
\;\in\;M_{2(n+1)}(\Aa^+)\;,
\qquad
V\,\in\,M_{2n}(\Aa^+)
\;.
\end{equation}
Then the quotient $K_0(\Aa)=\Vv_0(\Aa)\slash\sim_0$ becomes an abelian group with neutral element $0=[E_2]$ via
\begin{equation}
\label{eq-semigroup}
[V]\;+\;[V']\;=\;\left[
\begin{pmatrix}
V & 0 \\ 0 & V'
\end{pmatrix}
\right]
\;.
\end{equation}
By  \cite{BL,GS}, this definition of $K_0(\Aa)$ is equivalent  to the standard one which can be found {\it e.g.} in \cite{RLL,GVF}. The standard way to introduce the group $K_1(\Aa)$ is to set
$$
\Vv_1(\Aa)
\;=\;
\left\{
U\in \cup_{n\geq 1}M_{n}(\Aa^+)\;:\;
U^{-1}\,=\,U^*
\right\}
\;,
$$
and to define an equivalence relation $\sim_1$ by homotopy and $[U]=[\binom{U\;0}{0\;\one}]$. Then $K_1(\Aa) =\Vv_1(\Aa)/\sim_1$ with addition again defined by $[U]+[U']=[U\oplus U']$. If $\Aa$ is unital, one can work with $M_n(\Aa)$ instead of $M_n(\Aa^+)$ in $\Vv_1(\Aa)$, without changing the definition of $K_1(\Aa)$. 

\vspace{.2cm}

$K$-theory connects the $K$-groups of a given short exact sequence
\begin{equation}
\label{eq-ShortExact}
0 \;\rightarrow \;\Kk \;\rightarrow\; \Alge\;\overset{\pi}{\rightarrow} \;\Qq \;\rightarrow \;0 
\end{equation}
of C$^*$-algebras in an associated exact sequence of abelian groups. The main focus is here on one of the connecting maps, namely the index map $\Ind: K_1(\Qq)\to K_0(\Kk)$. While there is a standard definition of this map \cite{RLL,GVF}, let us bring it into a form convenient for our purposes, as in \cite{BL}.

\begin{proposi}
\label{prop-IndexMap}
Let the contraction $B\in M_n(\Alge^+)$ be a lift of a unitary $U\in M_n(\Qq^+)$, namely $\pi^+(B)=U$ where $\pi^+:\Alge^+\to\Qq^+ $ is the natural extension of $\pi$ in \eqref{eq-ShortExact}. Then
\begin{equation} 
\label{classic_boundary_of_unitary}
\Ind [U]
\;=\;[V]
\;,
\qquad
V\;=\;
\begin{pmatrix}
2BB^{*}-\one & 2B\sqrt{\one-B^{*}B}\\
2B^{*}\sqrt{\one-BB^{*}} & \one-2B^{*}B
\end{pmatrix}
\;.
\end{equation}
\end{proposi}

\noindent {\bf Proof. } First of all, let us note that indeed $V\in\Kk^+$ is a self-adjoint unitary $V\in\Kk^+$ with $s(V)\sim_0 E_{2n}$ because $\pi^+(2BB^*-\one)=\one$ and $\pi^+(\one-2B^*B)=-\one$, and $B^{*}\sqrt{\one-BB^{*}}=\sqrt{\one-B^*B}\,B^*$. The definition of $\Ind$ as given in \cite{RLL} uses a lift $W\in\Alge^+$ of $\diag(U,U^*)$ and is
\begin{equation}
\label{eq-IndIntermed}
\Ind [U]
\;=\;
\varphi_0
\left(
\left[W
\begin{pmatrix}
\one & 0 \\ 0 & 0
\end{pmatrix}
W^*\right]
\;-\;
\left[
\begin{pmatrix}
\one & 0 \\ 0 & 0
\end{pmatrix}
\right]
\right)
\;,
\end{equation}
where $\varphi_0$ is the map defined in \cite[Proposition~10]{GS} identifying the standard projection picture of $K_0(\Aa)$ from \cite{RLL} to \eqref{eq-V0def}. Due to \eqref{eq-equirel}, this eliminates the second summand in \eqref{eq-IndIntermed} and leads to 
%
$$
\Ind [U]
\;=\;
\left[
2\,W
\begin{pmatrix}
\one & 0 \\ 0 & 0
\end{pmatrix}
W^*
\,-\,\begin{pmatrix}
\one & 0 \\ 0 & \one
\end{pmatrix}
\right]
\;.
$$
Choosing
$$
W
\;=\;
\begin{pmatrix}
B &  -\sqrt{\one-BB^*} \\ \sqrt{\one-B^*B}  & B^*
\end{pmatrix}
\;,
$$
now concludes the proof.
\hfill $\Box$

\vspace{.2cm}

Under supplementary hypothesis, it is possible to choose further refined representatives of the image of the index map. 
The following version is tailored to analyze odd index pairings of the type $\Pi A\Pi+\one-\Pi$ where $A$ is unitary and $\Pi$ is a projection. It will be shown in Section~\ref{sec-winding} below how this abstract result can be applied to a concrete situation. 

\begin{theo}
\label{theo-abstract}
Suppose $0  \rightarrow \Kk \rightarrow \Alge \overset{\pi}{\rightarrow} \Qq \rightarrow 0 $
is a short exact sequence of $C^*$-algebras with $\Qq$ unital.  Suppose $A\in \Alge$ is unitary and that $\TapeP$ and $\TapeN$ are elements of $\Alge$ satisfying
$$
0\,\leq \,\TapeP, \TapeN \,\leq \,\one
\;,
\qquad
\TapeP\TapeN \,=\, 0
\;,
\qquad
 [\TapeP,A], \, [\TapeN,A] ,\, \TapeP^4 + \TapeN^4 - \one \,\in\, \Kk \;.
$$
This implies that
$$ 
U\; =\; \pi(\TapeP A\TapeP + \TapeN^2)
$$
is a unitary in $\Qq$. Let us introduce the hermitian operator $\WSL$ by
$$
\begin{pmatrix}
2\TapeP^4 + 2\TapeN^4 - \one & 
2(\one - \TapeP^4)^\frac{1}{4}\TapeP A\TapeP(\one - \TapeP^4)^\frac{1}{4} + \TapeN^2(\one - \TapeN^4)^\frac{1}{2} \\
2(\one - \TapeP^4)^\frac{1}{4}\TapeP A^*\TapeP(\one - \TapeP^4)^\frac{1}{4} + \TapeN^2(\one - \TapeN^4)^\frac{1}{2}  & \one - 2\TapeP^4 - 2\TapeN^4
\end{pmatrix} 
\,,
$$
and denote 
$$
\delta\;=\;\|[\TapeP^2,A] \|
\;.
$$
Then, with $\VSL$ given by \eqref{classic_boundary_of_unitary} with the lift $\TapeP A\TapeP+\TapeN^2$ of $U$,
$$
\|\VSL-\WSL\|
\;\leq\;
2\,\delta\,+\,4\,\delta^\frac{1}{4}
\;.
$$
If
\begin{equation} 
\label{eq-Condition}
\delta\; < \;0.0036\;,
\end{equation}
the operator $W$ is invertible and the boundary map in $K$-theory is
$$
\Ind\,[U]
\;=\;[\WSL_+  \,-\, \WSL_-] 
\;,
$$
where $\WSL_+$ and $\WSL_-$ are the positive and negative spectral projections of $\WSL$. 
\end{theo}

The key principle behind the proof is the following: one may perturb the matrix entries of $\VSL$ in \eqref{classic_boundary_of_unitary} without changing the associated $K_0$-class as long as the perturbation is sufficiently small so that the spectral gap of $\VSL$ does not close under the perturbation. An adequate measure of the size of the perturbation is the quantity $\delta$. For the formal proof, some technical preparations are needed. 

\begin{lemma} 
\label{lem:root_continuity}
If $\TapeP$ and $\TapeN$ are positive elements of a $C^*$-algebra and $0 \leq \alpha \leq 1$, then
$$
\| \TapeP^\alpha - \TapeN^\alpha \| 
\;\leq \;
\| \TapeP - \TapeN \|^\alpha 
$$
\end{lemma}

\noindent {\bf Proof.} 
This is proved in \cite{And} for matrices and the proof transposes to compact operators, but we could not find this result anywhere for the operator norm on infinite dimensional Hilbert spaces so we include a short proof. By \cite[Haagerup's Lemma]{Pede}, if $U$ is unitary and $Q \geq 0$, then
$$
\|[ U, Q^\alpha ] \| \;\leq \;\|[ U, Q ] \|^\alpha
\;.
$$
When applied to
$$
Q \;= \;\begin{pmatrix}
\TapeP & 0 \\
0 & \TapeN
\end{pmatrix}\;,
 \qquad
 U\; =\; \begin{pmatrix}
0 & \one \\
\one & 0
\end{pmatrix} 
\;,
$$
one finds
$$
\left \|
\begin{pmatrix}
0 & \TapeN^\alpha - \TapeP^\alpha \\
\TapeP^\alpha - \TapeN^\alpha & 0
\end{pmatrix}
\right\| 
\;\leq\;
\left \|
\begin{pmatrix}
0 & \TapeN - \TapeP \\
\TapeP - \TapeN & 0
\end{pmatrix}
\right\| ^\alpha
\;,
$$
which proves the lemma.
\hfill $\Box$

\begin{lemma} \label{lem:X_part}
Suppose $A$ is a unitary and $0 \leq \TapeP \leq \one$ in some unital $C^*$-algebra.  Then
$$
\left\| 2(\TapeP A\TapeP)(\TapeP A\TapeP)^* - \one - (2\TapeP^4 - \one)  \right\|
\;\leq\; 2\,\delta
\;,
$$
where $\delta = \| [\TapeP^2,A]\|$.
\end{lemma}

\noindent {\bf Proof.} 
This is basic: $\| (\TapeP A\TapeP)(\TapeP A\TapeP)^* - \TapeP^4 \| \,= \,\| \TapeP A\TapeP^2A^*\TapeP - \TapeP^3AA^*\TapeP \| \, \leq \,\| A\TapeP^2 - \TapeP^2A \|$.
\hfill $\Box$

\begin{lemma} \label{lem:A_part}

Suppose $A$ is a unitary and $0 \leq \TapeP \leq \one$ in some unital $C^*$-algebra.  Then
$$
\left\| 2(\TapeP A\TapeP)\sqrt{\one - (\TapeP A\TapeP)^*(\TapeP A \TapeP)} - 2\TapeP(\one-\TapeP^4)^\frac{1}{4}A\TapeP(\one-\TapeP^4)^\frac{1}{4}  \right\|
\;\leq\; 
4\,\delta^\frac{1}{4}
$$
where $\delta = \| [\TapeP^2,A]\|$.
\end{lemma}

\noindent {\bf Proof.} 
We notice 
$$
2(\TapeP A\TapeP)\sqrt{\one - (\TapeP A\TapeP)^*(\TapeP A\TapeP)} \;=\; 
2\left(\one - (\TapeP A\TapeP)(\TapeP A\TapeP)^* \right)^\frac{1}{4}\TapeP A\TapeP\left(\one - (\TapeP A\TapeP)^*(\TapeP A\TapeP) \right)^\frac{1}{4}
$$
giving us a more symmetric formula as a starting point. From the previous lemma, one finds
$$
\|  \one - (\TapeP A\TapeP)^*(\TapeP A\TapeP) - ( \one- \TapeP^4) \| \;\leq \;\delta
$$
so using Lemma~\ref{lem:root_continuity} we deduce
$$
\| \left(\one - (\TapeP A\TapeP)^*(\TapeP A\TapeP) \right)^\frac{1}{4} - \left(\one - \TapeP^4 \right)^\frac{1}{4} \| 
\;\leq \;\delta^\frac{1}{4}
$$
and, by symmetry,
$$
\| \left(\one - (\TapeP A\TapeP)(\TapeP A\TapeP)^* \right)^\frac{1}{4} - \left(\one - \TapeP^4 \right)^\frac{1}{4} \| 
\;\leq \;\delta^\frac{1}{4} .
$$
These imply
$$
\left\|
\left(\one - (\TapeP A\TapeP)(\TapeP A\TapeP)^* \right)^\frac{1}{4}\TapeP A\TapeP\left(\one - (\TapeP A\TapeP)^*(\TapeP A\TapeP) \right)^\frac{1}{4}
-\left(\one - \TapeP^4 \right)^\frac{1}{4}\TapeP A\TapeP\left(\one - \TapeP^4 \right)^\frac{1}{4}
\right\| 
\;\leq \;
2\,\delta^\frac{1}{4}
\;,
$$
concluding the proof. \hfill $\Box$

\vspace{.2cm}

\noindent {\bf Proof of Theorem~\ref{theo-abstract}.} We estimate the distance from $\WSL$ to the unitary $\VSL$ given by \eqref{classic_boundary_of_unitary} with the lift $\TapeP A\TapeP+\TapeN^2$ for $U$. Let us, just for the purpose of this proof, use the notation $A_\TapeP=\TapeP A\TapeP$. Then $V$ is given by
$$
\VSL = \begin{pmatrix}
2(A_\TapeP+\TapeN^2)(A_\TapeP+\TapeN^2)^*-\one & 2(A_\TapeP+\TapeN^2)\sqrt{\one-(A_\TapeP+\TapeN^2)^*(A_\TapeP+\TapeN^2)}\\
2(A_\TapeP+\TapeN^2)^*\sqrt{\one-(A_\TapeP+\TapeN^2)(A_\TapeP+\TapeN^2)^*} & \one-2(A_\TapeP+\TapeN^2)^*(A_\TapeP+\TapeN^2)
\end{pmatrix} 
.
$$
Notice that $\TapeP\TapeN=0$ implies that
$$
\VSL\;=\; \begin{pmatrix}
2(A_\TapeP)(A_\TapeP)^*-\one & 2(A_\TapeP)\sqrt{\one-(A_\TapeP)^*(A_\TapeP)}\\
2(A_\TapeP)^*\sqrt{\one-(A_\TapeP)(A_\TapeP)^*} & \one-2(A_\TapeP)^*(A_\TapeP)
\end{pmatrix} 
+ \begin{pmatrix}
2\TapeN^4-\one & 2\TapeN^2\sqrt{\one-\TapeN^4}\\
2\TapeN^2\sqrt{\one-\TapeN^4} & \one-2\TapeN^4
\end{pmatrix}
$$
and on the other hand
$$
\WSL =
\begin{pmatrix}
2\TapeP^4  - \one & 
\!\!\!\!\!
2(\one - \TapeP^4)^\frac{1}{4}A_\TapeP(\one - \TapeP^4)^\frac{1}{4} \\
2(\one - \TapeP^4)^\frac{1}{4}A^*_\TapeP(\one - \TapeP^4)^\frac{1}{4}   & \one - 2\TapeP^4
\end{pmatrix}
+ \begin{pmatrix}
2\TapeN^4-\one & 2\TapeN^2\sqrt{\one-\TapeN^4}\\
2\TapeN^2\sqrt{\one-\TapeN^4} & \one-2\TapeN^4
\end{pmatrix}.
$$
Setting
$$
A_{1,1} \;=\; 2A_\TapeP A_\TapeP^*-\one - (2\TapeP^4-1)
\;,
\qquad
A_{2,2} \;=\; 2A_\TapeP^*A_\TapeP-\one - (2\TapeP^4-\one)
\;,
$$
and 
$$
A_{1,2} \;=\; 2A_\TapeP\sqrt{\one-A_\TapeP^*A_\TapeP} - 2(\one - \TapeP^4)^\frac{1}{4}A_\TapeP(\one - \TapeP^4)^\frac{1}{4}\;,
$$
$$
A_{2,1} \;=\; 2A_\TapeP^*\sqrt{\one-A_\TapeP A_\TapeP^*} - 2(\one - \TapeP^4)^\frac{1}{4}A^*_\TapeP(\one - \TapeP^4)^\frac{1}{4}\;,
$$
one has
$$
\VSL-\WSL\;=\;
\begin{pmatrix}
A_{1,1} & 0 \\
0 & A_{2,2}
\end{pmatrix}
\,+\,
\begin{pmatrix}
0 & A_{1,2} \\
A_{2,1} & 0
\end{pmatrix}
\;,
$$
so that Lemmas~\ref{lem:X_part} and \ref{lem:A_part} imply
\begin{equation}
\label{eq-errorbound}
\|\VSL-\WSL\|
\;\leq\;
\max \{\|A_{1,1}\|,\|A_{2,2}\|\})\,+\,\max \{\|A_{1,2}\|,\|A_{2,1}\|\}
\;\leq\;
2\,\delta\,+\,4\,\delta^\frac{1}{4}
\;.
\end{equation}
As $\VSL$ is a hermitian unitary, any condition that forces $\delta + 2\delta^\frac{1}{4} < \frac{1}{2}$
will assure that the gap of $\WSL$ remains open.  Since $\delta + 2\delta^\frac{1}{4}$ is an increasing function and
$$
0.0036 + 2\times 0.0036 \,\approx\,  0.4935
$$
the condition $\delta < 0.0036$ will work.
\hfill $\Box$

\section{Application to the Hilbert space over the lattice $\ZM^d$}
\label{sec-winding}

Here we choose the algebras in the short exact sequence \eqref{eq-ShortExact} to be the bounded operators $\Alge=\Bb(\Hh)$ and compact operators $\Kk=\Kk(\Hh)$ on the separable Hilbert space $\Hh=\ell^2(\ZM^d)\otimes\CM^N$. Hence $\Qq$ is the associated Calkin algebra. The key idea is to choose $\TapeP=\TapePFunc(D)$ and $\TapeN=\TapeNFunc(D)$ in Theorem~\ref{theo-abstract} to be given in terms of two functions $\TapePFunc,\TapeNFunc:\RM\to [0,1]$ of the form
$$
p(x)\;=\;
\left\{
\begin{array}{cc}
0\;, & x\leq -\Radius\;, \\
p(x)
\;,
& 
|x|\leq\Radius\;,
\\
1 \;,& x\geq \Radius\;,
\end{array}
\right.
\qquad\quad
n(x)\;=\;
\left\{
\begin{array}{cc}
1\;, & x\leq -\Radius\;, \\
0 \;,& x> -\Radius\;,
\end{array}
\right.
$$
where $p$ is  supposed to be smooth and increasing. Then $\TapeP$ and $\TapeN=\TapeN^2$ are compact perturbations of $\Pi$ and $\one-\Pi$ respectively and therefore
$$
\Ind\big(\Pi A\Pi+(\one-\Pi)\big)
\;=\;
\Ind\big(\TapeP A\TapeP+\TapeN^2\big)
\;.
$$
We now want to apply Theorem~\ref{theo-abstract} to the r.h.s.. Indeed, $P$ and $N$ automatically satisfy many of the conditions in Theorem~\ref{theo-abstract}. In particularly, $[A,\TapeP],[A,\TapeN]\in \Kk$ follows from the fact that $A$ is local in the sense \eqref{eq-CommBound}, which implies that $\langle k|A|m\rangle\to 0$ as $|k-m|\to\infty$. Also $P^4+N^4-\one$ is compact and, what is crucial in the argument below, actually non-vanishing only on a ball of size $\Radius$.   What is missing is merely to check that the estimate \eqref{eq-Condition} holds. This is connected to a judicious choice of $p$ and depends on the following fact.

\begin{proposi}[Theorem~3.2.32 in \cite{BR}] 
\label{prop-BR}
For differentiable $f:\RM\to\RM$ and Fourier transform defined with normalization factor $\frac{1}{2\pi}$,
$$
\|[f(D),A]\|
\;\leq\;
\|\widehat{f'}\|_{L^1(\RM)}\;\|[D,A]\|
\;.
$$
\end{proposi}

\begin{lemma} 
\label{lem-tapering0}
There exists an even differentiable function $G_\Radius:\RM\to[0,1]$ with 
$$
G_\Radius(x)\;=\;
\left\{
\begin{array}{cc}
0 \;,& \;\;\;|x|\geq \Radius\;, 
\\
1\;, & \;\;\;|x|\leq\frac{\Radius}{2}\;,
\end{array}
\right.
$$
such that $\|\widehat{G'_\Radius}\|_{L^1(\RM)}\leq \frac{8}{\Radius}$.
\end{lemma}

\noindent {\bf Proof.} (The choice below goes back to the late Uffe Haagerup.) Let us first consider $\Radius=1$ and then rescale later on. The construction starts by setting $f'(x)=\max\{0,1-|x|\}$ which has integral $1$.  Integrating, one finds for $|x|\leq 1$
$$
f(x)\;=\;
\left\{
\begin{array}{cc}
\tfrac{1}{2}(1+x)^2\;, & \;\;\;\;x\in[-1,0]\;,
\\
1-\tfrac{1}{2}(1-x)^2\;, & \;\;\;x\in[0,1]\;.
\end{array}
\right.
$$
On the other hand, one calculates that $\widehat{f'}(p)=\frac{1-\cos(p)}{\pi p^2}$ and consequently $\|\widehat{f'}\|_{L^1(\RM)}=1$. Now set $G_1(x)=f(4x+3)-f(4x-3)$ for which $\|\widehat{G'_1}\|_{L^1(\RM)}\leq 8$. Then $G_\Radius(x)=G_1(\frac{x}{\Radius})$ has all the desired properties.
\hfill $\Box$

\vspace{.2cm}

Associated to $G_\Radius$ will be a function $F_\Radius$ increasing from $0$ to $1$ in $[-\Radius,\Radius]$ and satisfying $G_\Radius^4=4F_\Radius(1-F_\Radius)$, a relation which stems from Theorem~\ref{theo-abstract} when choosing $P=F_\Radius(D)$, see below. Thus $F_\Radius(x)=\frac{1}{2}\big(1+\sgn(x)(1+G_\Radius(x)^4)^{\frac{1}{2}}\big)\in[0,1]$. Of course, one could construct a function $F_\Radius$ increasing from $0$ to $1$ in $[-\Radius,\Radius]$ more directly by the technique of Lemma~\ref{lem-tapering0} and this actually would improve the quantitative estimate in Proposition~\ref{prop-tuneddown}. However, even such an improved estimate would not be sufficient for the proof of Theorem~\ref{theo-main}, and the main point of Proposition~\ref{prop-tuneddown} is rather that it already allows to calculate the index pairing as the signature of a finite matrix. On the other hand, the function $G_\Radius$ will be our best choice in the proof of Theorem~\ref{theo-main} later on, and lead to the quantitative estimate stated there.

\begin{proposi}
\label{prop-tuneddown}
Suppose that $A\in \Bb(\Hh)$ is unitary and that $\Radius>0$ is such that
\begin{equation}
\label{eq-cond2}
\|[D,A]\|\;<\;\frac{\Radius}{32}\; (0.0036)^4
\;.
\end{equation}
Then $\Pi A\Pi+(\one-\Pi)$ is Fredholm and its index is equal to the (well-defined!) signature of the hermitian matrix
\begin{equation}
\label{eq-BottLocal}
\SL(F_\Radius,{\TapeG})
\;=\;
\begin{pmatrix}
2F_\Radius -\one_\Radius & \TapeG A \TapeG\\ \TapeG A^*\TapeG & -2F_\Radius+\one_\Radius
\end{pmatrix}
\;,
\end{equation}
where $G_\Radius=G_\Radius(D_\Radius)$ and $F_\Radius=F_\Radius(D_\Radius)$, and $D_\Radius$ and $\one_\Radius$ are the restrictions of $D$ and $\one$ to $\ell^2(\DM_\Radius)\otimes\CM^N$ over the discrete disc $\DM_\Radius=\{x\in\ZM^d\,:\,\|x\|\leq \Radius\}$. Here $\|x\|$ denotes the euclidean norm.
\end{proposi}

\noindent {\bf Proof. } Using $P=F_\Radius(D)^{\frac{1}{4}}$, Haagerup's inequality and then Proposition~\ref{prop-BR} shows
\begin{align*}
\|[P^2,A]\|
& 
\;=\;
\|[F_\Radius(D)^{\frac{1}{2}},A]\|
\;\leq\;
\|[F_\Radius(D),A]\|^{\frac{1}{2}}
\;\leq\;
\|[G_\Radius(D)^4,A]\|^{\frac{1}{4}}
\\
& 
\;\leq\;
\Big(4\,\|[G_\Radius(D),A]\|\Big)^{\frac{1}{4}}
\;\leq\;
\Big(\frac{32}{\Radius}\;\|[D,A]\|\Big)^{\frac{1}{4}}
\;<\;
0.0036
\;,
\end{align*}
by hypothesis \eqref{eq-cond2}. Therefore \eqref{eq-Condition} holds and the $K$-theoretic index map of $\TapeP A\TapeP+\TapeN^2$ is given in Theorem~\ref{theo-abstract} in terms of the hermitian $\WSL$. Now the restriction to $\DM_\Radius^c=\ZM^d\setminus \DM_\Radius$ of $\TapeP^4+\TapeN^4$ is the identity $\one_\Radius^c$ and $\TapeN^2(\one - \TapeN^4)^\frac{1}{2}=0$. Furthermore, $\one - \TapeP^4=(\one - \TapeP^4)\one_\Radius=\one_\Radius - \TapeP^4_\Radius$. Thus the operator $\WSL$ has a direct sum representation on $\big(\ell^2(\DM_\Radius)\oplus \ell^2(\DM_\Radius^c)\big)\otimes\CM^{2N}$  given by
$$
\begin{pmatrix}
2\TapeP_\Radius^4  - \one_\Radius & 2\TapeP_\Radius(\one_\Radius - \TapeP_\Radius^4)^\frac{1}{4}A\TapeP_\Radius(\one_\Radius - \TapeP_\Radius^4)^\frac{1}{4}\\
2\TapeP_\Radius(\one_\Radius - \TapeP_\Radius^4)^\frac{1}{4}A^*\TapeP_\Radius(\one_\Radius - \TapeP_\Radius^4)^\frac{1}{4}  & \one_\Radius - 2\TapeP_\Radius^4
\end{pmatrix} 
\oplus
\begin{pmatrix}
\one_\Radius^c & 0 \\ 0 & -\one_\Radius^c
\end{pmatrix}
\;.
$$
Therefore we have expressed the index in terms of the signature of a finite matrix:
$$
\Ind\big(\Pi A\Pi+(\one-\Pi)\big)
\;=\;
\Sig
\begin{pmatrix}
2\TapeP_\Radius^4  - \one_\Radius & 
\!\!\!\!
2\TapeP_\Radius(\one_\Radius - \TapeP_\Radius^4)^{\frac{1}{4}}A\TapeP_\Radius(\one_\Radius - \TapeP_\Radius^4)^{\frac{1}{4}}\\
2\TapeP_\Radius(\one_\Radius - \TapeP_\Radius^4)^{\frac{1}{4}}A^*\TapeP_\Radius(\one_\Radius - \TapeP_\Radius^4)^{\frac{1}{4}}  & \one_\Radius - 2\TapeP_\Radius^4
\end{pmatrix} 
\;.
$$
Replacing $P_\Radius=F_\Radius(D_\Radius)^{\frac{1}{4}}$ now implies the claim because $\sqrt{2}\TapeP_\Radius(\one_\Radius - \TapeP_\Radius^4)^{\frac{1}{4}}=G_\Radius$.
\hfill $\Box$

\section{Quantitative estimate on stabilization of signature}
\label{sec-estimate}

In this section we complete the proof of Theorem~\ref{theo-main}. This requires us  to deform the functions $F_\Radius$ and $G_\Radius$ in Proposition~\ref{prop-tuneddown} in such a manner that \eqref{eq-BottLocal} becomes the finite volume spectral localizer $\SL_{\Tune,\Radius}$ defined in \eqref{eq-DefBott2}. During the deformation the finite matrix has to remain invertible so that the signature does not change. Moreover, the following quantitative estimate on the size of the gap of $\SL_{\Tune,\Radius}$  assures that the signature in Theorem~\ref{theo-main} is well-defined.

\begin{theo}
\label{theo-SigEstimate}
Let $A$ be an invertible, bounded operator on $\Hh=\ell^2(\ZM^d)\otimes\CM^N$. With $g=\|A^{-1}\|^{-1}$, suppose that  $\Tune>0$ and $\Radius<\infty$ are such that the bounds \eqref{eq-MainCond1} and \eqref{eq-MainCond2} in Theorem~\ref{theo-main} hold. Then 
\begin{equation}
\label{eq-GapBound}
\SL_{\Tune,\Radius}^2\;\geq\;\frac{g^2}{2}
\;.
\end{equation}
\end{theo}

\noindent {\bf Proof. } Both the aim described in the introduction to this section and Theorem~\ref{theo-SigEstimate} can be attained by the same technique. Therefore, as a preparation for the proof of Theorem~\ref{theo-main}, let us consider the more general case of 
$$
\SL_{\Tune}(F,G) 
\;=\;
\begin{pmatrix}
\Tune\,\Radius\, (2F-\one_\Radius)  & G A_\Radius G \\ G A_\Radius^* G &  -\,\Tune\,\Radius\, (2F-\one_\Radius)
\end{pmatrix}
\;,
$$
where $F=F(D_\Radius)$ and $G=G(D_\Radius)$ is built from smooth functions $F,G:\RM\to [0,1]$ given by
\begin{equation}
\label{eq-Homotopy}
F
\;=\;
\lambda F^L\,+\,(1-\lambda)F_\Radius
\;,
\qquad
G
\;=\;
\lambda G^L\,+\,(1-\lambda)G_\Radius
\;,
\end{equation}
where $\lambda\in[0,1]$ and 
$$
F^L(x)
\;=\;
\frac{1}{2}\Big(1+\frac{x}{\Radius}\Big)
\;,
\qquad
G^L(x)\;=\;1\;.
$$ 
For the latter two functions, one has $\SL_{\Tune}(F^L,G^L)= \SL_{\Tune,\Radius}$, and on the other hand $\SL_{\frac{1}{\Radius}}(F_\Radius,{\TapeG})$ is the matrix $\SL(F_\Radius,{\TapeG})$ in \eqref{eq-BottLocal} in Proposition~\ref{prop-tuneddown}. Therefore $\SL_{\Tune}(F,G)$ allows to connect the matrix in Proposition~\ref{prop-tuneddown} to the spectral localizer by a smooth two-parameter homotopy in $\lambda$ and $\Tune$. What has to be assured in the following is that $\SL_{\Tune}(F,G) $ remains invertible throughout. The existence of a gap of $\SL_{\Tune}(F,G) $ will follow from a lower bound on
\begin{align*}
\SL_{\Tune}(F,G)^2
\;=\; &
\Tune^2\,\Radius^2\,
\begin{pmatrix}
(2F-\one_\Radius)^2 &  0 \\ 0 & (2F-\one_\Radius)^2  
\end{pmatrix}
\,+\, 
\begin{pmatrix}
GA_\Radius G^2A_\Radius^*G &  0 \\ 0 & GA_\Radius^*G^2A_\Radius G
\end{pmatrix}
\\
&
\;\,+\,2\,\Tune\,\Radius\,
\begin{pmatrix}
0 &  [F,GA_\Radius G] \\ [GA_\Radius^*G,F] & 0
\end{pmatrix}
\;.
\end{align*}
The first two summands are non-negative and will be shown to combine to a strictly positive operator. The last term is an error which has to be controlled. 

\vspace{.2cm}

Let us begin with a lower bound on the second summand. Due to $G\geq G_\Radius$,
\begin{align*}
GA_\Radius^*G^2A_\Radius G
& \;=\;
GA^*G^2AG
\\
& \;\geq\;
 GA^* \TapeG^2 AG
\\
& \;=\;
 G\TapeG A^*A\TapeG G
\,+\,
G\big([A^*,\TapeG]\TapeG A+\TapeG A^*[\TapeG,A]\big)G
\\
& \;\geq\;
g^2\,G\,\TapeG^2\,G
\,+\,
G\,\big([A^*,\TapeG]\TapeG A+\TapeG A^*[\TapeG,A]\big)\,G
\;,
\end{align*}
and similarly for $GA_\Radius G^2A_\Radius^* G$. The error term here is bounded using Proposition~\ref{prop-BR} with the function from Lemma~\ref{lem-tapering0}:
$$
\|G\,\big([A^*,\TapeG]\TapeG A+\TapeG A^*[\TapeG,A]\big)\,G\|
\;\leq\;2\,\|A\|\,\|G\|\,\|G_\Radius\|\,\|[\TapeG,A]\|
\;\leq\;2\,\|A\|\,\frac{8}{\Radius}\,\|[D,A]\|
\;.
$$
An estimate on the third summand is obtained from
\begin{align*}
\|
[F,GA_\Radius G]
\|
&
\;\leq\;
\|G\|^2\,
\|
[F,A]
\|
\\
&
\;\leq\;
\lambda
\,
\|
[F^L,A]
\|
\;+\;
(1-\lambda)\,
\|[F_\Radius,A]\|
\\
&
\;=\;
\frac{\lambda}{2\Radius}\,
\|[D,A]\|
\,+\,
(1-\lambda)\,
\|[F_\Radius,A]\|
\,.
\end{align*}
Just replacing the first summand and dealing with the error terms as in \eqref{eq-errorbound}, one obtains by combining all the above and suppressing the $2\times 2$ matrix degree 
\begin{align}
\SL_{\Tune}(F,G)^2
\;\geq\;
& \Tune^2\,\Radius^2\,(2F-\one_\Radius)^2
\,+\,
g^2\,G^2\,\TapeG^2
\nonumber
\\
& 
-\,
\frac{16}{\Radius}\,
\|A\|\,\|[D,A]\|
\,-\, 
\Tune\,\lambda\,
\|[D,A]\|
\,+\,
\Tune\,\Radius\,(1-\lambda)\,
\|[F_\Radius,A]\|
\;.
\label{eq-GenGapEst}
\end{align}
For the proof of Theorem~\ref{theo-SigEstimate}, let us now choose $\lambda=1$ so that $G=G^L=\one_\Radius$ and $F=F^L$. The first term will then be bounded by
$$
 \Tune^2\,\Radius^2\,(2F^L-\one_\Radius)^2
 \;=\;
\Tune^2\, (D_\Radius)^2
\;\geq\;
g^2\,(\one_\Radius-G_\Radius^2)
\;,
$$
as $ \Tune\Radius\geq 2 g$ by \eqref{eq-MainCond2}. Indeed, then the bound  holds on $\DM_\Radius\setminus \DM_{\frac{\Radius}{2}}$  as $\one_\Radius-G_\Radius^2 \leq \one_\Radius$, while it holds trivially on $\DM_{\frac{\Radius}{2}}$ where $\one_\Radius-G_\Radius^2=0$. Replacing in \eqref{eq-GenGapEst} it follows that
$$
\SL_{\Radius,\Tune}^2
\;\geq\;
g^2\,\one_\Radius
\,-\,
\Big(
\frac{16}{\Radius}\,
\|A\|\,
\,+\, 
\Tune\,
\Big)
\|[D,A]\|
\;\geq\;
g^2\,\one_\Radius
\,-\,9\,
\frac{\Tune}{g}\,
\|A\|\,
\|[D,A]\|
\;,
$$
where the inequalities $\|A\|\geq 1$ and $g\leq 1$ were used, as well as \eqref{eq-MainCond2} in the form $\frac{1}{\Radius}\leq \frac{\Tune}{2g}$. Now \eqref{eq-MainCond1} leads to $\SL_{\Radius,\Tune}^2\geq\frac{g^2}{2}\,\one_\Radius$.
\hfill $\Box$

\vspace{.2cm}

\noindent {\bf Proof} of Theorem~\ref{theo-main}: As already explained above, the first step consists in showing that the path \eqref{eq-Homotopy}  is within the invertible matrices for all  $\lambda$. This is done using the lower bound \eqref{eq-GenGapEst} and will first be done for a unitary $A$ only. Then the error term $\|[F_\Radius,A]\|$ in \eqref{eq-GenGapEst} can be bounded by Haagerup's equality exactly as in the proof of Proposition~\ref{prop-tuneddown}. The other main variation on the above argument is how to estimate $2F-\one_\Radius$ from below. This will be based on some analysis of the functions $F^L$ and $F_\Radius$. One can check that
$$
(2F^L(x)-1) \;\geq\;c_F (2 F_\Radius(x)-1)
$$
holds for $x\geq 0$ and $c_F=\frac{3}{4}$. As the functions
$$
2F(x)-1
\;=\;
\lambda (2F^L(x)-1)\;+\;
(1-\lambda)(2 F_\Radius(x)-1)
\;.
$$
are odd and all three positive for $x\geq 0$, one thus has
\begin{align*}
(2F(x)-1)^2
& 
\;\geq\;
\lambda^2 (2F^L(x)-1)^2\;+\;
(1-\lambda)^2(2 F_\Radius(x)-1)^2
\\
&
\;\geq\;\frac{1}{4}\,c_F^2\, (2 F_\Radius(x)-1)^2
\\
&
\;=\;\frac{1}{4}\,c_F^2\, (1-G_\Radius(x)^4)
\;.
\end{align*}
This bound holds uniformly in $\lambda\in[0,1]$. Replacing in \eqref{eq-GenGapEst}  shows
\begin{align*}
\SL_{\Tune}(F,G)^2
\;\geq\;
& \Tune^2\,\Radius^2\,\frac{1}{4}\,c_F^2\, (\one_\Radius-G_\Radius^4)
\,+\,
g^2\,\TapeG^4
\\
& 
-\,
\frac{16}{\Radius}\,
\|A\|\,\|[D,A]\|
\,-\, 
\Tune\,\lambda\,
\|[D,A]\|
\,+\,
\Tune\,\Radius\,(1-\lambda)\,
\Big(\frac{32}{\Radius}\;\|[D,A]\|\Big)^{\frac{1}{2}}
\;.
\end{align*}
Now let us choose $\Tune=\frac{1}{\Radius}$. Then
$$
\SL_{\frac{1}{\Radius}}(F,G)^2
\;\geq\;
\frac{1}{4}\,c_F^2\, \one_\Radius
\;-\;
\frac{C}{\Radius^{\frac{1}{2}}}\,\max\{
\|[D,A]\|,\|[D,A]\|^{\frac{1}{2}}
\}
$$
for some constant $C$. In particular, for $\Radius$ large enough, $\SL_{\frac{1}{\Radius}}(F,G)$ remains for all $\lambda\in[0,1]$.  As already pointed out, for $\lambda=0$ the matrix $\SL_{\frac{1}{\Radius}}(F,G)$ is  that given in  \eqref{eq-BottLocal} , which by Proposition~\ref{prop-tuneddown} has a signature equal to  $2\,\Ind\big(\Pi A\Pi+(\one-\Pi)\big)$. By homotopy this is also true for $\lambda=1$, for which due to the choice \eqref{eq-Homotopy} of the path leads to $\SL_{\frac{1}{\Radius}}(F,G)$ being equal $\SL_{\frac{1}{\Radius},\Radius}$. Now in a second step, one can change the parameters $\Tune$ and $\Radius$ in $\SL_{\Tune,\Radius}$. As long as the bounds \eqref{eq-MainCond1} and \eqref{eq-MainCond2} hold Theorem~\ref{theo-SigEstimate} implies that the signature of $\SL_{\Tune,\Radius}$ does change. (Strictly speaking, $\SL_{\frac{1}{\Radius},\Radius}$ falls out of the range of parameters of Theorem~\ref{theo-main} if the gap $g$ is larger than $\frac{1}{2}$, but this can always be circumvented by decreasing $g$ as a parameter.) This concludes the proof of Theorem~\ref{theo-main} for a unitary $A$. If $A$ is merely invertible, one first uses the polar decomposition to deform it by $t\in[0,1]\mapsto A|A|^{-t}$ into a unitary. Along this path the index does not change. For the unitary the estimates \eqref{eq-MainCond1} and \eqref{eq-MainCond2} may be worse, but they still allow the above argument to run. At the end, one can deform back to $A$ and note that Theorem~\ref{theo-SigEstimate} did {\it not} use the unitarity of $A$.
\hfill $\Box$


\section{The $\eta$-invariant of the spectral localizer}
\label{sec-EtaBott}

\begin{proposi}
\label{prop-EtaExist}
Let $d=1$. Suppose that \eqref{eq-CommBound} holds, namely that $\|[D,A]\|<\infty$, and that $\SL_\Tune$ defined in \eqref{eq-DefBott} is invertible. Then $\SL_\Tune$  has a well-defined $\eta$-invariant.
\end{proposi}

The proof is a combination of nowadays standard techniques, {\it e.g.} \cite{JLO,GeS,Get,CP}.

\vspace{.2cm}

\noindent {\bf Proof.} We will use the representation \eqref{eq-etadef2} of the $\eta$-function $\eta_s(\SL_\Tune)$ in terms of the heat kernel of $\SL_\Tune^2$ and split it into $\eta_s(\SL_\Tune)=\eta'_s(\SL_\Tune)+\eta''_s(\SL_\Tune)$ with
$$
\eta'_s(\SL_\Tune)
\;=\;
\frac{1}{\Gamma(\tfrac{s+1}{2})}
\int_0^1 dt\;t^{\tfrac{s-1}{2}}\;
\Tr\big(\SL_\Tune\,e^{-t\SL_\Tune^2}\big)
\;,
\qquad
\eta''_s(\SL_\Tune)
\;=\;
\frac{1}{\Gamma(\tfrac{s+1}{2})}
\int_1^\infty dt\;t^{\tfrac{s-1}{2}}\;
\Tr\big(\SL_\Tune\,e^{-t\SL_\Tune^2}\big)
\;.
$$
Hence estimates on the trace norm of the heat kernel are needed. They will be obtained by a perturbative argument. Let us write 
$$
\SL_\Tune^2\;=\;
\Delta+\Perturb
\;,
$$ 
where $\Delta=\Tune^2 (\DiracP)^2$ and
$$
\Perturb 
\;=\;
\begin{pmatrix}
AA^* & \Tune [D,A] \\ \Tune [D,A]^* & A^*A
\end{pmatrix}
\;,
$$
Let us note that $\Delta\geq 0$ and that, by hypothesis, $\Perturb $ is a bounded operator which is viewed as a perturbation. Furthermore, $\Delta$ is even w.r.t. the grading  where $J=\binom{\one\;\;0}{0\,-\one}$, namely $\Delta J=J\Delta$, and $H$ is odd as $HJ=-JH$. Replacing DuHamel's formula 
\begin{equation}
\label{eq-DuHamel}
e^{-t\SL_\Tune^2}
\;=\;
e^{-t\Delta}
\,+\,
t\,\int^1_0dr\,e^{-(1-r)t\Delta}\Perturb e^{-rt \SL_\Tune^2}
\;,
\end{equation}
formally into \eqref{eq-etadef2} leads to
\begin{equation}
\label{eq-DuHamelRep}
\eta'_s(\SL_\Tune)
\;=\;
\frac{1}{\Gamma(\tfrac{s+1}{2})}
\int_0^1 dt\;t^{\tfrac{s-1}{2}}\;
\left(
\Tr\big(\SL_\Tune\,e^{-t\Delta}\big)
\,+\,
t \int^1_0dr\,\Tr\big(\SL_\Tune e^{-(1-r)t\Delta}\Perturb e^{-rt\SL_\Tune^2}\big)
\right)
\;.
\end{equation}
Of course, we have to show that all traces and integrals exist, but this will be achieved below. Then the aim is to show that the limit $s\to0$ exists. For this we will use that $e^{-t\Delta}$ is traceclass for all $t>0$. The crucial fact for this is that the first summand vanishes because, with $\SL_\Tune=\Tune \DiracP +H$ as in \eqref{eq-DefBott},
$$
\Tr\big(\SL_\Tune\,e^{-t\Delta}\big)
\;=\;
\Tune \,\Tr\big(\DiracP \,e^{-t\Delta}\big)
\;+\;
\Tr\big(H\,e^{-t\Delta}\big)
\;,
$$
and $\Tr(\DiracP  e^{-t\Delta})=\Tr(\DiracP  e^{-t \Tune^2 \DiracP ^2})=0$ due to the symmetry of the  spectrum of $\DiracP =D\otimes\sigma_3$, and furthermore $\Tr(He^{-t\Delta})=0$ using the cyclicity of the trace as well as the fact that $H$ is odd while $e^{-t\Delta}$ is even w.r.t.\ $J$. Hence it only remains to bound the second summand in \eqref{eq-DuHamelRep} and show that it remains bounded as $s\to 0$. For that purpose, let us again decompose $\SL_\Tune=\Tune \DiracP +H$ and focus on the contribution containing $\DiracP $. By Cauchy-Schwarz,
\begin{align*}
\big|\Tr\big(\Tune \DiracP  e^{-(1-r)t\Delta }\Perturb e^{-rt\SL_\Tune^2}\big)\big|^2
&\;\leq\;
\Tr\big(\Delta e^{-2(1-r)t\Delta})\,\Tr(\Perturb ^*\Perturb e^{-2rt\SL_\Tune^2}\big)
\\
&\;\leq\;
\|\Perturb \|^2\,\Tr\big(\Delta e^{-2(1-r)t\Delta}\big)\,\Tr\big(e^{-2rt\SL_\Tune^2}\big)
\;.
\end{align*}
As the the spacial dimension is $d=1$, the factor $\Tr(\Delta e^{-2(1-r)t\Delta})$ can be bounded by the derivative of a Gaussian integral, namely a constant times $(t-tr)^{-\frac{3}{2}}$. To bound $\Tr(e^{-2rt\SL_\Tune^2})$, let us expand the heat kernel into a norm-convergent Dyson series by using iteratively DuHamel's formula \eqref{eq-DuHamel}:
$$
e^{-t\SL_\Tune^2}
\,=\,
e^{-t\Delta}
\,+\,
\sum_{n=1}^\infty t^n\!\int^1_0\!\!dr_1\int^{r_1}_0\!\!dr_2\cdots \int^{r_{n-1}}_0\!\!dr_n\,e^{-(1-r_1)t\Delta}\Perturb e^{-(r_2-r_1)t \Delta}\Perturb \cdots 
\Perturb e^{-r_nt\Delta}
\;.
$$
Now the first trace norm $\|e^{-t\Delta}\|_1 =\Tr(e^{-t\Delta})$ can for $d=1$ be bounded above by a Gaussian integral, and hence by a constant times $t^{-\frac{1}{2}}$. If $\DiracP $ were invertible and thus $\Delta$ strictly positive, then $\|e^{-t\Delta}\|_1$ would have exponential decay for large $t$ by the arguments below, but this will actually not be needed. Rather, using the multiple H\"older inequality for the inverse exponent $(1-r_1)+(r_2-r_3)+\ldots+r_n=1$, the positivity of $e^{-t\Delta}$ and the norm estimate $\|AB\|_p\leq\|A\|\,\|B\|_p$ for the Schatten norms leads to
$$
\|e^{-t\SL_\Tune^2}\|_1
\;\leq\;
\|e^{-t\Delta}\|_1
\;+\;
\sum_{n=1}^\infty t^n\!\int^1_0\!\!dr_1\int^{r_1}_0\!\!dr_2\cdots \int^{r_{n-1}}_0\!\!dr_n\,
\Big(\prod^n_{j=0}
\|e^{-(r_j-r_{j+1})t\Delta}\|_{\frac{1}{r_j-r_{j+1}}}\Big)
\|\Perturb \|^{n}
$$
where we set $r_0=1$ and $r_{n+1}=0$. Hence summing up the series
\begin{equation}
\label{eq-TraceNorm}
\|e^{-t\SL_\Tune^2}\|_1
\;\leq\;
e^{t\|\Perturb \|}\, \|e^{-t\Delta}\|_1
\;\leq\;
C\,e^{t\|\Perturb \|}\,t^{-\frac{1}{2}}
\;.
\end{equation}
Consequently, with a constant $C'$ depending also on $\epsilon$ and $\|\Perturb \|$,
$$
\big|\Tr\big(\Tune \DiracP  e^{-(t-r)\Delta }\Perturb e^{-r\SL_\Tune^2}\big)\big|
\;\leq\;
C'\,
(t-r)^{-\frac{3}{4}}\,r^{-\frac{1}{4}}
\;.
$$
As $\Tr(H e^{-(t-r)\Delta }\Perturb e^{-r\SL_\Tune^2})$ is bounded by the same expression, replacing in the above shows
$$
\eta'_s(\SL_\Tune)
\;\leq\;
\frac{1}{|\Gamma(\tfrac{s+1}{2})|}
\int_0^1 dt\;t^{\tfrac{\Re e(s)-1}{2}}\;
\int^t_0dr\,C''\,\|\Perturb \|\,(t-r)^{-\frac{3}{4}}\,r^{-\frac{1}{4}}
\;<\;\infty
\;,
$$
as long as $\Re e(s)>-1$, so in particular for $s=0$.

\vspace{.1cm}

For $\eta''_s(\SL_\Tune)$ and hence large $t$, the estimate \eqref{eq-TraceNorm} is of little help. It has to be boosted by using the gap of $\SL_\Tune$. Suppose that $\SL_\Tune^2\geq \epsilon$, a lower bound that can be given by \eqref{eq-GapBound} as $\epsilon=\frac{g^2}{2}$. Then, for any $\alpha\in (0,1)$, by Cauchy-Schwarz
\begin{align*}
\Tr\big(\SL_\Tune\,e^{-t\SL_\Tune^2}\big)^2
& \;\leq\;
\Tr\big(\SL_\Tune^2\,e^{-2\alpha t\SL_\Tune^2}\big)\;
\Tr\big(e^{-2(1-\alpha) t\SL_\Tune^2}\big)
\\
& \;\leq\;
(2\alpha t)^{-1}\;\Tr\big(e^{-\alpha t\SL_\Tune^2}\big)
\;\| e^{-2(1-2\alpha) t\SL_\Tune^2} \|\;
\Tr\big(e^{-2\alpha t\SL_\Tune^2}\big)
\;,
\end{align*}
where the bound $x e^{-x t}\leq t^{-1}e^{-\frac{x t}{2}}$ for $x,t>0$ was used. Hence with \eqref{eq-TraceNorm}
$$
\Tr\big(\SL_\Tune\,e^{-t\SL_\Tune^2}\big)
\;\leq\;
(2\alpha t)^{-\frac{1}{2}}\,e^{-(1-2\alpha) t\epsilon}\,\Tr\big(e^{-\alpha t\SL_\Tune^2}\big)
\;\leq\;
(2\alpha t)^{-\frac{1}{2}}\,e^{-(1-2\alpha) t\epsilon}\,
e^{\alpha t\|\Perturb \|}\,C\,(\alpha t)^{-\frac{1}{2}}
\;.
$$
Choosing $\alpha \leq \frac{1}{4}\min\{1,\frac{\epsilon}{\|\Perturb \|}\}$, one infers that for some constant $C'''$ depending on $\epsilon$
\begin{equation}
\label{eq-TraceNorm2}
\Tr\big(\SL_\Tune\,e^{-t\SL_\Tune^2}\big)
\;\leq\;
C'''\,e^{-\frac{t\epsilon}{4}}
\;.
\end{equation}
Hence also $\eta''_s(\SL_\Tune)$ is bounded, actually for all $s$.
\hfill $\Box$

\vspace{.2cm}

Combining Proposition~\ref{prop-EtaExist} with Theorem~\ref{theo-SigEstimate} taken in the limit $\Radius\to\infty$ leads to:

\begin{coro}
\label{coro-EtaExist}
Let $A$ be an invertible, bounded operator on $\Hh=\ell^2(\ZM)\otimes\CM^N$. Suppose that  $\Tune>0$ is sufficiently small such that \eqref{eq-MainCond1} holds. Then $\eta(\SL_\Tune)$ exists.
\end{coro}

The following result connects the $\eta$-invariant to the spectral flow $\SF$ of a certain path of unbounded selfajoint operators with compact resolvent. Given such a path, the spectral flow is intuitively defined as the spectrum passing $0$ from left to right along the path, minus the spectrum passing from right to left. We refer to \cite{Phi,CP} for a careful definition of the spectral flow. That there is a general connection between $\eta$-invariants and spectral flow is already proved in \cite{Get,CP}, and the following result is merely a corollary of these papers.

\begin{theo}
\label{theo-etaSF}
Let $d=1$.  Suppose that $\SL_\Tune=\Tune \DiracP +H$ has a well-defined $\eta$-invariant. Consider the path $\lambda\in[0,1]\mapsto \SL_\Tune(\lambda)=\Tune \DiracP +\lambda H$ of selfadjoint operators with compact resolvents.  Then
$$
\eta(\SL_\Tune)
\;=\;
2\,\SF\big(\lambda\in[0,1]\mapsto \SL_\Tune(\lambda)\big)
\;.
$$
\end{theo}

\noindent {\bf Proof.} Let us first give an intuitive argument. According to Proposition~\ref{prop-EtaExist}, $\SL_\Tune(\lambda)$ has a well-defined $\eta$-invariant as long as it is invertible. Precisely when it is not invertible, there is a crossing of an eigenvalue by $0$. Each such crossing modifies the $\eta$-invariant by $\pm 2$, pending on whether the eigenvalue moves from left to right or from right to left. Thus twice the spectral flow gives  $\eta(\SL_\Tune(1))-\eta(\SL_\Tune(0))$. But $\eta(\SL_\Tune(1))=\eta(\SL_\Tune)$ because $\SL_\Tune(1)=\SL_\Tune$, and $\eta(\SL_\Tune(0))=\eta(\Tune \DiracP )=0$ because the spectrum of $\DiracP $ is symmetric, see the proof of Proposition~\ref{prop-EtaExist}. 

\vspace{.1cm}

The more formal proof uses Theorem~2.6 of \cite{Get} or, equivalently, Corollary~8.10 of \cite{CP} for the I$_\infty$ case.   Indeed, the regularized $\eta$-invariants $\eta_\epsilon(\SL_\Tune(1))$ and $\eta_\epsilon(\SL_\Tune(0))$ converge in the limit $\epsilon\downarrow 0$ to the $\eta$-invariants by Proposition~\ref{prop-EtaExist} (here the index $\epsilon$ denotes as in \cite{Get,CP} a cut-off on the integration domain $[0,\infty)$ to $[\epsilon,\infty)$ and {\it not} the parameter $s$). Then the above mentioned results conclude the proof provided that
\begin{equation}
\label{eq-SupTermVan}
\lim_{\epsilon\downarrow 0}\,\sqrt{\tfrac{\epsilon}{\pi}}\,\int^1_0 d\lambda\;\Tr\big(\partial_\lambda \SL_\Tune(\lambda)\,e^{-\epsilon\,\SL_\Tune(\lambda)^2}\big)
\;=\;
0
\;.
\end{equation}
As $\partial_\lambda \SL_\Tune(\lambda)=H$,  one hence needs to control $\Tr(He^{-\epsilon \SL_\Tune(\lambda)^2})$. This will be done uniformly in $\lambda$, so let us drop the argument $\lambda$ (or absorb it in $H$). Let us simply replace DuHamel's formula \eqref{eq-DuHamel} to deduce

\vspace{-.3cm}

$$
\Tr\big(He^{-\epsilon \SL_\Tune^2}\big)
\;=\;
\Tr\big(He^{-\epsilon \Delta}\big)
\,+\,
\epsilon\,\int^1_0dr\,\Tr\big(He^{-(1-r)\epsilon \Delta}\Perturb e^{-r\epsilon \SL_\Tune^2}\big)
\;.
$$
As above, the first summand $\Tr(H\,e^{-t\Delta})$ vanishes because $H$ is odd and $\Delta$ is even w.r.t. $J$. Hence using Cauchy-Schwarz and then \eqref{eq-TraceNorm}
\begin{align*}
\Tr\big(He^{-\epsilon \SL_\Tune^2}\big)
& \;\leq\;
\epsilon\,\|H\|\,\|\Perturb \| 
\int^1_0dr\;
\Big(
\|e^{-2(1-r)\epsilon \Delta}\|_1\,\|e^{-r\epsilon \SL_\Tune^2}\|_1
\Big)^{\frac{1}{2}}
\\
& \;\leq\;
\epsilon\,\|H\|\,\|\Perturb \| 
\int^1_0dr\;
\Big(
C \big(2(1-r)\epsilon\big)^{-\frac{1}{2}}
\,C\,
e^{r\epsilon\|\Perturb \|}\,
\big(r\epsilon\big)^{-\frac{1}{2}}
\Big)^{\frac{1}{2}}
\;.
\end{align*}
This readily implies \eqref{eq-SupTermVan}, and thus concludes the formal proof.
\hfill $\Box$

\vspace{.2cm}

As it is well-known \cite{Phi,CP} that the spectral flow has certain homotopy invariance properties, one can deduce the following corollary by combining Corollary~\ref{coro-EtaExist} (essentially Proposition~\ref{prop-EtaExist} and Theorem~\ref{theo-SigEstimate}) with  Theorem~\ref{theo-etaSF}. Let us stress that the argument leading to it does not use the results of Sections~\ref{sec-ImageInd} and \ref{sec-winding}.

\begin{coro}
\label{coro-EtaHomotopy}
Suppose that $\lambda\in[0,1] \mapsto A(\lambda)$ is a continuous path of invertibles on $\ell^2(\ZM)$ and that $\Tune$ is such that the locality bound \eqref{eq-MainCond1} holds uniformly in $\lambda$. If $\SL_\Tune(\lambda)$ is defined from $A(\lambda)$ by \eqref{eq-DefBott}, the associated $\eta$-invariants $\eta(\SL_\Tune(\lambda))$ are well-defined and constant in $\lambda$.
\end{coro}

To conclude this section, we will show how Theorem~\ref{theo-etaSF} can be used to calculate the $\eta$-invariant and thus also the finite volume signature. This provides an alternative, purely analytic proof of Theorem~\ref{theo-main} which does not depend on the $K$-theoretic arguments of Sections~\ref{sec-ImageInd} and \ref{sec-winding}. However, we only treat the case of dimension $d=1$. The argument also shows that \eqref{eq-MainCond1} is close to optimal.  

\vspace{.2cm}

\noindent {\bf Sketch of an alternative proof} of Theorem~\ref{theo-main} for $d=1$: By the homotopy invariance of the spectral flow, it is sufficient to prove the connection between Fredholm index and $\eta$-invariant for the $n$-fold right shift operators $A=S^n$, for all $n\in\ZM$. These operators form a set of representatives for each $K_1$-class (for the Banach algebra of operators on $\ell^2(\ZM)$ with bounded $[D,A]=[X,A]$), and the indices of the associated Fredholm operators $\Pi S^n\Pi+\one-P$ are equal to $n$ and thus also in bijection with $\ZM$. Hence we are lead to study the spectral flow of the path
$$
\lambda\in[0,1]\;\mapsto\; \SL_\Tune(\lambda)
\;=\;
\begin{pmatrix}
\Tune\,X & \lambda\,S^n \\ \lambda\,(S^*)^n & -\,\Tune\,X
\end{pmatrix}
\;,
$$
and to verify that this spectral flow is $n$. The spectrum of $\SL_\Tune(\lambda)$ can be determined explicitly by solving the eigenvalue equation for a spectral parameter $b$:
$$
\SL_\Tune(\lambda)\begin{pmatrix} \phi \\ \psi\end{pmatrix} 
\;=\;
\begin{pmatrix}
\Tune\,X \phi\,+\, \lambda\,S^n \psi \\ \lambda\,(S^*)^n \phi\, -\,\Tune\,X\psi
\end{pmatrix}
\;=\;
b\begin{pmatrix} \phi \\ \psi\end{pmatrix} 
\;.
$$
Multiplying the second equation by $\lambda$ leads to $\lambda^2(S^*)^n\phi=\lambda (b+\Tune X)\psi$, and replacing this into
the first equation multiplied by by $(b+\Tune X)(S^*)^n$  leads to
$$
\Tune(b+\Tune X)(S^*)^nX\phi\;+\;\lambda^2(S^*)^n\phi\;=\;b(b+\Tune X)(S^*)^n\phi
\;.
$$
Applying $S^n$ and using $SXS^*=X-1$ shows
$$
\Tune(b+\Tune (X-n))X\phi\;+\;\lambda^2\phi\;=\;b(b+\Tune (X-n))\phi
\;.
$$
Hence $\phi$ has to be diagonal in the eigenbasis of $X$, namely it is a state $\phi=|k\rangle$ localized at site $k\in\ZM$. This is possible provided
$$
\Tune(b+\Tune (k-n))k\;+\;\lambda^2\;=\;b(b+\Tune (k-n))
\quad
\Longleftrightarrow
\quad
b_\pm(k)\;=\;\tfrac{\Tune n}{2}\,\pm\left((\tfrac{\Tune n}{2}-\Tune k)^2+\lambda^2\right)^{\frac{1}{2}}
\;.
$$
Hence the spectrum for fixed $n$ is
$$
\sigma(\SL_\Tune(\lambda))
\;=\;
\left\{
\tfrac{\Tune }{2}\Big(n\,\pm\left((n-2k)^2+\tfrac{4\lambda^2}{\Tune^2}\right)^{\frac{1}{2}}\Big)\;:\;k\in\ZM
\right\}
\;.
$$
For say $n>0$, only eigenvalues $b_-(k)$ with $k\in (0,n]$  can cross $0$, precisely when
$$
n^2\,-\,(n-2k)^2\;=\;\tfrac{4\lambda^2}{\Tune^2}
\;.
$$
This always happens for some $\lambda\in[0,1]$ when $\frac{2}{\Tune}\geq n$. But $\|[D,A]\|=\|[X,S^n]\|=n$, so that the condition reads $\frac{2}{\Tune}\geq \|[D,A]\|$ which holds due to \eqref{eq-MainCond1}, as $g=1$ and $\|A\|=1$ in the present situation. As all these $n$ eigenvalues $b_-(1),\ldots,b_-(n)$ move from left to right, the spectral flow is $+n$ and the claim is checked.
\hfill $\Box$


\vspace{.5cm}

\noindent {\bf Acknowledgements:} The authors thank Martin Zirnbauer and the ESI in Vienna for inviting them to the program ``Topological Phases of Quantum Matter'' in 2014. That was where the project was initiated. We also thank the FAU for support in form of a Visiting Professorship for T.~L. in 2016, as well as the Simons Foundation and the DFG for financial support.  The first named author would like to express his gratitude to the mathematics department at FAU for their kindness and hospitality.

\vspace{1cm}


\end{document}